\documentclass[12pt]{article}

\pdfoutput=1

\usepackage{amsmath}
\usepackage{amssymb}
\usepackage{amsfonts}

\usepackage{graphicx}
\usepackage{color}

\newcommand{\vect}[1]{\textit{\textbf{#1}}}

\newcommand{\me}{\mathrm{e}}
\newcommand{\dif}{\mathrm{d}}
\newcommand{\Omvec}{\mathbf{\Omega}}

\title{
A Generalized Linear Transport Model for Spatially-Correlated Stochastic Media
}

\author{
Anthony B. Davis \\ (corresponding author) \\
\& \\
Feng Xu \\
Jet Propulsion Laboratory \\
California Institute of Technology \\
4800 Oak Grove Drive (Mail Stop 233-200) \\
Pasadena, CA 91109, USA \\
Anthony.B.Davis@jpl.nasa.gov \\
http://science.jpl.nasa.gov/people/ADavis/
}

\date{\today}

\begin{document}

\maketitle

Article to appear in \emph{Journal of Computational and Theoretical Transport}, formerly known as \emph{Statistical Physics and Transport Theory} (until 2013), as part of a Special Issue dedicated to Ken Case's legacy for the {\bf 23rd International Conference on Transport Theory (ICTT23)}, Santa Fe, NM, Sept 15-19, 2013.

\begin{abstract}
We formulate a new model for transport in stochastic media with long-range spatial correlations where exponential attenuation (controlling the propagation part of the transport) becomes power law.  Direct transmission over optical distance $\tau(s)$, for fixed physical distance $s$, thus becomes $(1+\tau(s)/a)^{-a}$, with standard exponential decay recovered when $a\to\infty$.  Atmospheric turbulence phenomenology for fluctuating optical properties rationalizes this switch.  Foundational equations for this generalized transport model are stated in integral form for $d=1,2,3$ spatial dimensions.  A deterministic numerical solution is developed in $d=1$ using Markov Chain formalism, verified with Monte Carlo, and used to investigate internal radiation fields.  Standard two-stream theory, where diffusion is exact, is recovered when $a=\infty$.  Differential diffusion equations are not presently known when $a<\infty$, nor is the integro-differential form of the generalized transport equation.  Monte Carlo simulations are performed in $d=2$, as a model for transport on random surfaces, to explore scaling behavior of transmittance $T$ when transport optical thickness $\tau_\text{t} \gg 1$.  Random walk theory correctly predicts $T \propto \tau_\text{t}^{-\min\{1,a/2\}}$ in the absence of absorption.  Finally, single scattering theory in $d=3$ highlights the model's violation of angular reciprocity when $a<\infty$, a desirable property at least in atmospheric applications.  This violation is traced back to a key trait of generalized transport theory, namely, that we must distinguish more carefully between two kinds of propagation: one that ends in a virtual or actual detection, the other in a transition from one position to another in the medium.
\end{abstract}

\section{Introduction: Motivation \& Outline}
\label{s:intro}

All natural optical media are to some extent variable in space, often in such a complex way that they are best represented with statistics.  In nuclear engineering, there is increasing interest in pebble-bed reactors where the core is made of many small spheres that contain both fuel and moderator material.  In contrast with classic reactor designs, their detailed 3D geometry (i.e., how the spheres stack) is quite random.  Earth's cloudy atmosphere is another instance of a very clumpy 3D optical medium.  These are just two examples from vastly different disciplines where a good theory for stochastic transport would be a valuable asset.

Broadly speaking, three kinds of model have been proposed to account for unresolved spatial variability in a transport medium.
\begin{itemize}
\item
The most natural approach is ``homogenization'' where one seeks \emph{effective} material properties that can be used in the solution of a transport problem for a uniform medium, but would make an accurate prediction of the behavior of the heterogeneous stochastic medium.  The homogenized material properties will depend on statistical quantities (means, variances, correlations, etc.) that characterize the stochastic medium of interest.  Examples for the cloudy atmosphere are in \cite{DavisEtAl1990,Cahalan1994,Cairns2000}.
\item
An alternative is to develop new transport equations to solve either analytically or numerically.  Examples for the cloudy atmosphere are in \cite{AvasteVainikko1974,Stephens1988b,Davis2006}.  Interestingly, the early paper by Avaste and Vyanikko \cite{AvasteVainikko1974} proposes a binary mixture model that has a long and ongoing history of application to nuclear engineering, going at least back to the seminal papers by Levermore, Pomraning et al. \cite{LevermorePomraning1986,LevermoreEtAl1988}.  This approach is at least conceptually more difficult than the previous one since new methods must be found to solve the new transport equations.
\item
A third approach, of intermediate complexity, is to linearly combine the answers of a number of computations for uniform media in order to approximate the answer for the spatially heterogeneous stochastic medium.  Examples of application to the cloudy atmosphere are in \cite{Mullamaa1972,RonnholmEtAl1980,StephensEtAl_TTSP91,CahalanEtAl1994a,Barker1996a,Barker_etal08}.
\end{itemize}
In our experience, homogenization will work well for weaker kinds of variability and/or higher tolerance for error.  A model derived using the second approach, such as the one proposed in the following pages, should be more broadly applicable.  Models of the 3rd kind can be competitive, largely due to their straightforward implementation.

In the following, we will primarily keep clouds and atmospheric optics in mind, but the generalized transport model we propose may prove to be more broadly applicable.  Accordingly, we will talk about radiative transfer (RT) and RT equations (RTEs), but the entirety of this work can be thought of as transport theory as defined by the linear Boltzmann equation.

\paragraph{Outline:}
The remainder of this article is organized as follows.
Section~\ref{s:dDim_sRTE} introduces our notations and states the standard RTE and boundary conditions for homogeneous---or random but ``homogenized''---plane-parallel media in $d$ spatial dimensions ($d=1,2,3$).
Section~\ref{s:dDim_gRTE} introduces our ansatz leading to a new class of generalized RTEs in integral form with power-law transmission laws.  Therein, we first see how non-exponential transmission laws arise from the statistics of stochastic media, with an emphasis on the role of spatial correlations, as exemplified by the Earth's turbulent and cloudy atmosphere.
In Section~\ref{s:2stream_MarCh}, the $d=1$ case gets special attention.  In the framework of standard RT, it is formally identical to the well-known two-stream model.  Turning to generalized RT, we derive ab initio a deterministic numerical solution in $d=1$, and use it to investigate internal radiation fields.  The new generalized RT solver is based on Markov chain formalism, traditionally a tool for random walk theory (including its application to Monte Carlo methods in transport).  A technical Appendix details the computational methodology used in the Markov chain code.
Section~\ref{s:DiffusionLimits} revisits the behavior of diffuse transmission in the absence of absorption for standard and generalized RT in the diffusion limit (i.e., asymptotically large transport optical depth).  New numerical experiments in $d=2$ validate the theoretical prediction based on self-similar L\'evy flights.  This reduced dimensionality is easier to comprehend graphically, and also may have applications in transport phenomena on random surfaces.
In Section~\ref{s:ReciprocityViolation}, we use the single scattering limit in $d=3$ to show that generalized RT is not reciprocal under a switch of sources and detectors.  This violation of angular reciprocity is in fact observed in the Earth's cloudy atmosphere---the original motivation and application of the generalized RT model.
In the final Section~\ref{s:concl}, we present our conclusions and an outlook on practical applications of our theoretical and computational advances, including a connection with recent work atmospheric spectroscopy \cite{ConleyCollins2011}.

\section{Standard Radiative Transport in $d$ Spatial Dimensions}
\label{s:dDim_sRTE}

\subsection{RTE for Homogeneous---or Homogenized---Media in Integro-Differential Form}

Let $I(z,\Omvec)$ denote the steady-state radiance field at level $z$ in a \emph{uniform} $d$-dimensional plane-parallel optical medium of thickness $H$,
\begin{equation}
\text{M}_d(H) = \{ \vect{x}\in\mathbb{R}^d, 0<z<H \},
\end{equation}
propagating in direction $\Omvec$ on the $d$-dimensional sphere,
\begin{equation}
\Xi_d = \{ \Omvec\in\mathbb{R}^d, \|\Omvec\|=1 \}.
\end{equation}
$I(z,\Omvec)$ has physical units of radiant power per unit of $d$-dimensional ``area'' per $d$-dimensional ``solid angle.''  Table~\ref{t:Definitions} gives explicit definitions of $\vect{x}$, $\Omvec$, and other properties introduced further on for $d = 1,2,3$.

Denoting the extinction coefficient (expressed in m$^{-1}$) by $\sigma$, $I(z,\Omvec)$ is a function of exactly $d$ variables that verifies the linear transport equation
\begin{equation}
\left[ \Omega_z\frac{\dif\;}{\dif z} + \sigma \right]
I(z,\Omvec) = S(z,\Omvec) + q(z,\Omvec),
\label{e:nD_RTE_uniform}
\end{equation}
where $S(z,\Omvec)$ is the (unknown) source \emph{function} for multiple scattering and $q(\vect{x},\Omvec)$ is the (specified) source \emph{term}.  These quantities have the physical units of [$I$] further divided by a unit of length, hence radiant power per unit of $d$-dimensional ``volume,'' instead of ``area.''  Specifically, we have
\begin{equation}
S(z,\Omvec) = \sigma_\text{s}\int\limits_{\Xi_d}
p(\Omvec^\prime\cdot\Omvec)I(z,\Omvec^\prime)\dif\Omvec^\prime,
\label{e:nD_RTE_SourceFunction}
\end{equation}
where $p(\Omvec^\prime\cdot\Omvec)$ is the phase function (PF) in units of inverse $d$-dimensional solid angle, which we assume is only a function of the scattering angle $\theta_\text{s} = \cos^{-1}\Omvec^\prime\cdot\Omvec$.  As an important example, we have listed in Table~\ref{t:Definitions} values for the PF when scattering is isotropic.  The quantity $\sigma_\text{s}$, appearing in (\ref{e:nD_RTE_SourceFunction}), is the scattering coefficient in m$^{-1}$.  Combining (\ref{e:nD_RTE_uniform}) and (\ref{e:nD_RTE_SourceFunction}) leads to the RTE in $d$ dimensions in standard \emph{integro-differential} form.

A popular approach for modeling RT in stochastic media is to use ``homogenized'' optical properties $\sigma$, $\sigma_\text{s}$ and $p(\Omvec^\prime\cdot\Omvec)$.  This means that, rather than simple averages over the $d$-dimensional spatial variability of actual optical properties, an \emph{effective} value is taken that somehow captures the average impact of the spatial fluctuations on $I(z,\Omvec)$, itself a spatial average radiance field.  The effective optical properties will depend on a subset of their respective means, variances, possibly higher-order moments, auto-correlations, cross-correlations, and so on.

Apart from previously mentioned physics-based homogenization techniques in RT for the cloudy atmosphere \cite{DavisEtAl1990,Cahalan1994,Cairns2000}, rigorous mathematical methods have been brought to bear on this still challenging problem; see, e.g., \cite{Allaire1992,DumasGolse2000,BalJing2010}.  However, these studies focus on highly oscillatory optical media.  Such high-frequency (``noisy'') stochastic media were independently investigated by Davis and Mineev-Weinstein \cite{DavisMineev11} that are predicated on power-law (scaling) statistics.  They used averaging methods akin to those described further on (in \S\ref{s:nonExpTran}), but with the necessary modifications to account for noise-like spatial variability.  Specifically, these authors assumed media where the extinction coefficient fluctuations have a wavenumber spectrum
\begin{equation}
E_\sigma(k) \sim k^{-\beta},
\label{e:PowerLaw_EnergySpectum}
\end{equation}
over a broad range of scales (i.e., $1/k$) that overlaps with radiatively relevant ones, including $H$ and the mean free path (defined rigorously further on for variable media).  They found that in cases of white- or blue-noise media ($\beta \le 0$), homogenization will likely work, being enabled by approximately exponential mean transmission laws.  Otherwise, that is, in cases of pink- or red-noise media ($0 < \beta \le 1$), it will not work since exponential decay is a poor approximation to the mean transmission law.  Media with $\beta > 1$ are not noise-like---they have a stochastic continuity property---and are discussed in \S\ref{s:nonExpTran}.

In the present study, we are exclusively interested in the response of uniform or stochastic media to irradiation from an external source.  If this source is collimated (highly concentrated into a single direction $\Omvec_0$, with $\Omega_{0z} > 0$), then we can take
\begin{equation}
q(z,\Omvec) = F_0 \exp(-\sigma z/\mu_0) \sigma_\text{s} p(\Omvec_0\cdot\Omvec)
\label{e:nD_RTE_SourceTerm}
\end{equation}
in the uniform case, where $F_0$ (in W/m$^{d-1}$) is its uniform areal density.  We also introduce here
\begin{equation*}
\mu_0 = \cos\theta_0 = \Omega_{z0}.
\end{equation*}
Note that we have oriented the $z$-axis positively in the direction of the incoming flow of solar radiation, as is customary in atmospheric optics.  The meaning of each factor in (\ref{e:nD_RTE_SourceTerm}) is clear: the incoming flux $F_0$ at $z = 0$ is attenuated exponentially (Beer's law) along the oblique path to level $z$ where it is scattered with probability $\sigma_\text{s}$ per unit of path length and, more specifically, into direction $\Omvec$ according to the PF value for $\theta_\text{s} = \cos^{-1}\Omvec_0\cdot\Omvec$.  In this case, $I(z,\Omvec)$ is the diffuse radiation (i.e., scattered once or more).

The appropriate boundary conditions (BCs) for the diffuse radiance that obeys (\ref{e:nD_RTE_uniform})--(\ref{e:nD_RTE_SourceTerm}) will express that none is coming in from the top of the medium, $I(0,\Omvec) = 0$ for $\Omega_z > 0$.  At the lower ($z = H$) boundary, we will take
\begin{equation}
I(H,\Omvec) = F_-(H) / c_d,
\label{e:nD_RTE_lowerBC_Lambertian}
\end{equation}
where
\begin{equation}
F_-(H) = \rho \, F_+(H),
\label{e:rho_Lambertian}
\end{equation}
for all $\Omega_z < 0$, where $\rho$ is the albedo of the partially ($0 < \rho < 1$) or totally ($\rho = 1$) reflective surface; we have also introduced the downwelling (subscript ``$+$'') and upwelling (subscript ``$-$'') hemispherical fluxes
\begin{equation}
\begin{array}{l}
F_+(z) = \int\limits_{\Omega_z>0} \Omega_z I(z,\Omvec)\dif\Omvec
       + \mu_0 F_0 \me^{-\sigma z/\mu_0}, \\
F_-(z) = \int\limits_{\Omega_z<0}|\Omega_z|I(z,\Omvec)\dif\Omvec.
\end{array}
\label{e:hemispherical_fluxes}
\end{equation}
This surface reflectivity model is, for simplicity, Lambertian (isotropically reflective), and the numerical constant $c_d = \int_{\Omega_z>0}\Omega_z\dif\Omvec$ is given in Table~\ref{t:Definitions} for $d = 1,2,3$.  Naturally, we will also consider a black (purely absorbing) surface in (\ref{e:nD_RTE_lowerBC_Lambertian}) by setting $\rho = 0$.

Alternatively, we can view $I(z,\Omvec)$ as total (uncollided \emph{and} once or more scattered) radiance, and assume $q(z,\Omvec) \equiv 0$ inside M$_d(H)$.  Radiation sources will then be represented in the expression of boundary conditions (BCs).  The upper ($z = 0$) BC expresses either diffuse or collimated incoming radiation.  In the former case, we have
\begin{equation}
I(0,\Omvec) = F_0 / c_d,
\label{e:nD_RTE_upperBC_diffuse}
\end{equation}
for any $\Omvec$ with $\Omega_z > 0$.  In the latter case, we have
\begin{equation}
I(0,\Omvec) = F_0 \delta(\Omvec-\Omvec_0),
\label{e:nD_RTE_upperBC_collimated}
\end{equation}
for $\Omega_z > 0$.  To reconcile (\ref{e:nD_RTE_SourceTerm}) with the above BC, we notice that
\begin{equation}
I_0(z,\Omvec) = F_0 \exp(-\sigma z/\mu_0) \delta(\Omvec-\Omvec_0)
\label{e:uncollided}
\end{equation}
is the solution of the ODE in (\ref{e:nD_RTE_uniform}) when the r.-h. side vanishes identically (no internal sources, nor scattering), and we use (\ref{e:nD_RTE_upperBC_collimated}) as the initial condition.  This uncollided radiance becomes the source of diffuse radiation immediately after scattering, hence its role in (\ref{e:nD_RTE_SourceTerm}).

In (\ref{e:uncollided}), $s = z/\mu_0$ is simply the oblique path covered by the radiation in the medium from its source at $z = s = 0$ to the location where it is detected, or scattered, or absorbed, or even escapes the medium ($s \ge H/\mu_0$).  From the well-known properties of the exponential probability distribution, this makes the mean free path (MFP) $\ell$ between emission, scattering or absorption events equal to the the e-folding distance $1/\sigma$.

Quantities of particular interest in many applications, including atmospheric remote sensing, are radiances at the boundaries that describe out-going radiation: $I(0,\Omvec)$ with $\Omega_z \le 0$; $I(H,\Omvec)$ with $\Omega_z \ge 0$.  Normalized (outgoing, hemispherical) boundary fluxes,
\begin{eqnarray}
\label{e:reflectivity}
R &=& \frac{F_-(0)}{\mu_0 F_0}, \\
\label{e:transmittivity}
T &=& \frac{F_+(H)}{\mu_0 F_0},
\end{eqnarray}
are also of interest, particularly, in radiation energy budget computations.  In (\ref{e:reflectivity})--(\ref{e:transmittivity}), the denominator is in fact $F_+(0)$ from (\ref{e:hemispherical_fluxes}).  Therefore, for the diffuse illumination pattern in (\ref{e:nD_RTE_upperBC_diffuse}), we only need to divide by $F_0$.

Finally, a convenient non-dimensional representation of out-going radiances, at least at the upper boundary, uses the ``bidirectional reflection factor'' (BRF) form:
\begin{equation}
I_\text{BRF}(\Omvec) = \frac{c_d I(0,\Omvec)}{\mu_0 F_0},
\label{e:BRF_form}
\end{equation}
for $\mu < 0$.  This is the ``effective'' albedo $\rho$ of the medium, i.e., as defined in (\ref{e:nD_RTE_lowerBC_Lambertian})--(\ref{e:rho_Lambertian}), but with $z=0$ rather than $z=H$, knowing $I(0,\Omvec)$ and hence $F_+(0) = \mu_0 F_0$.
Unlike the optical property $\rho$ in (\ref{e:rho_Lambertian}) and the radiative response $R$ in (\ref{e:reflectivity}), $I_\text{BRF}(\Omvec)$ is not restricted by energy conservation to the interval $[0,1]$.

\begin{table}[ht]
\caption{{\bf Definitions for }$d = 1,2,3$}
\label{t:Definitions}
\begin{center}
\begin{tabular}{|l||c|c|c|}
\hline
$d$ & 1 & 2 & 3 \\
\hline\hline
$\vect{x}$ & $z$ & $(x,z)^\text{T}$ & $(x,y,z)^\text{T}$ \\
\hline
$\dif\vect{x}$ & $\dif z$ & $\dif x\dif z$ & $\dif x\dif y\dif z$ \\
\hline
$\Omvec$   & $\pm 1$ & $(\sin\theta,\cos\theta)^\text{T}$ & $(\sin\theta\cos\phi,\sin\theta\sin\phi,\cos\theta)^\text{T}$ \\
\hline
$\dif\Omvec$ & n/a$^\dag$ & $\dif\theta$ & $\dif\cos\theta\dif\phi$ \\
\hline
$c_d$ in (\ref{e:nD_RTE_lowerBC_Lambertian}), (\ref{e:nD_RTE_upperBC_diffuse}) & 1 & 2 & $\pi$ \\
\hline
[$F_0$] in (\ref{e:nD_RTE_SourceTerm}), (\ref{e:nD_RTE_upperBC_diffuse})--(\ref{e:BRF_form}) & W & W/m & W/m$^2$ \\
\hline
[$I$] & W & W/m/rad & W/m$^2$/sr \\
\hline
[$S$] = [$q$] & W/m & W/m$^2$/rad & W/m$^3$/sr \\
\hline
$p_{d,\text{iso}} = p_0(\mu_\text{s})$ & 1/2 [-] & 1/2$\pi$ [rad$^{-1}$] & 1/4$\pi$ [sr$^{-1}$] \\
\hline
$p_g(\mu_\text{s})$ & $\frac{1+g\mu_\text{s}}{2}$ & $\left(\frac{1}{2\pi}\right)\frac{1-g^2}{1+g^2-2g\mu_\text{s}}$ & $\left(\frac{1}{4\pi}\right)\frac{1-g^2}{(1+g^2-2g\mu_\text{s})^{3/2}}$ \cite{HG_41}\\
\hline
$\chi_d$ in (\ref{e:dD_diffusion_BCs}) & 1 & $\pi$/4 & 2/3 \\
\hline
\end{tabular} \\
\end{center}
{\footnotesize $\null^\dag$In $d=1$, angular integrals become sums over the up ($\mu = -1$) and down ($\mu = +1$) directions, or only downward in (\ref{e:nD_RTE_lowerBC_Lambertian}).} \\
{\footnotesize N.B. In all cases, we use $\mu_\text{s} = \cos\theta_\text{s}$ to denote $\Omvec\cdot\Omvec^\prime$, the scalar product of the ``before'' and ``after'' scattering direction vectors.}
\end{table}

Actually, in the familiar $d = 3$ dimensions, all of the above is known as ``1D'' RT theory since only the spatial dimensions with any form of variability count.  If $\sigma$, $\sigma_\text{s}$ and $p(\cdot)$ depend on $z$, it is still 1D RT.  One can even remove from further consideration the former quantity by adopting the standard change of variables, $z \mapsto \tau = \int_0^z \sigma(z^\prime) \dif z^\prime$.  In this case, $z \mapsto \tau = \sigma z$ (depth in units of MFP $\ell = 1/\sigma$), then (\ref{e:nD_RTE_uniform})--(\ref{e:nD_RTE_SourceFunction}) become
\begin{equation}
\left[ \mu\frac{\dif\;}{\dif\tau} + 1 \right]
I(\tau,\Omvec) = \omega \int\limits_{\Xi_d}
p(\Omvec^\prime\cdot\Omvec)I(\tau,\Omvec^\prime)\dif\Omvec^\prime + q(\tau,\Omvec),
\label{e:1D_RTE}
\end{equation}
where $\mu$ denotes $\Omega_z$ ($=\cos\theta$ if $d>1$) and $\omega = \sigma_\text{s}/\sigma$ is the single scattering albedo (SSA).  We have assumed that $\omega$ and $p(\cdot)$ are independent of $z$, hence of $\tau$, for simplicity as well as consistency with the notion of a homogenized optical medium.

Another important non-dimensional property is the total optical thickness of the medium M$_3(H)$, namely, $\tau^\star = \sigma H = H/\ell$.  BCs for (\ref{e:1D_RTE}) are expressed as in (\ref{e:nD_RTE_lowerBC_Lambertian})--(\ref{e:nD_RTE_upperBC_collimated}) but at $\tau = 0,\tau^\star$.

Finally, we adopt the Henyey--Greenstein (H--G) PF model $p_g(\mu_\text{s})$ expressed in the penultimate row of Table~\ref{t:Definitions}.  Its sole parameter is the asymmetry factor $g = \int_{\Xi_d}\Omvec^\prime\cdot\Omvec p(\Omvec^\prime\cdot\Omvec)\dif\Omvec$.  The whole 1D RT problem is then determined entirely by the choice of four quantities, $\{\omega,g;\tau^\star;\rho\}$, plus $\mu_0$ if $d > 1$.

\subsection{Integral Forms of the $d$-Dimensional RTE}

Henceforth, we take $q(\tau,\Omvec) \equiv 0$ in (\ref{e:nD_RTE_uniform}) and, consequently, $I(\tau,\Omvec)$ is total (uncollided and scattered) radiation and the upper BC is (\ref{e:nD_RTE_upperBC_collimated}).  We will also assume in the remainder that $\rho = 0$ in the lower BC, cf. (\ref{e:nD_RTE_lowerBC_Lambertian})--(\ref{e:rho_Lambertian}), which then becomes simply $I(\tau^\star,\Omvec) = 0$ for $\mu < 0$.  These assumptions are not essential to our goal of generalizing RT theory to account for spatial heterogeneity with long-range correlations, but they do simplify many of the following expressions that are key to the discussion.

Now suppose that we somehow know $S(\tau,\Omvec)$ in (\ref{e:nD_RTE_uniform}), with $q(\tau,\Omvec) \equiv 0$.  It is then straightforward to compute $I(\tau,\Omvec)$ everywhere.  We simply use upwind integration or ``sweep:''
\begin{equation}
I(\tau,\Omvec) =
\left\{ \begin{array}{ll}
\int\limits_0^\tau S(\tau^\prime,\Omvec)\me^{-(\tau-\tau^\prime)/\mu}\frac{\dif\tau^\prime}{\mu}
+ I(0,\Omvec)\me^{-\tau/\mu},
& \text{if }\mu > 0, \\
\int\limits_\tau^{\tau^\star}
S(\tau^\prime,\Omvec)\me^{-(\tau^\prime-\tau)/|\mu|}\frac{\dif\tau^\prime}{|\mu|}
+ I(\tau^\star,\Omvec)\me^{-(\tau^\star-\tau)/|\mu|},
& \text{otherwise,}
\end{array} \right.
\label{e:S_sweep}
\end{equation}
where the boundary contributions are specified by the BCs.  When these BCs express an incoming collimated beam at $\tau = 0$, cf. (\ref{e:nD_RTE_upperBC_collimated}), and an absorbing surface at $\tau = \tau^\star$, cf. (\ref{e:nD_RTE_lowerBC_Lambertian})--(\ref{e:rho_Lambertian}) with $\rho = 0$, this simplifies to
\begin{equation}
I(\tau,\Omvec) =
\left\{ \begin{array}{ll}
\int\limits_0^\tau S(\tau^\prime,\Omvec)\me^{-(\tau-\tau^\prime)/\mu}\frac{\dif\tau^\prime}{\mu}
+ I_0(\tau,\Omvec),
& \text{if }\mu > 0, \\
\int\limits_\tau^{\tau^\star}
S(\tau^\prime,\Omvec)\me^{-(\tau^\prime-\tau)/|\mu|}\frac{\dif\tau^\prime}{|\mu|},
& \text{otherwise,}
\end{array} \right.
\label{e:simpleS_sweep}
\end{equation}
where $I_0(\tau,\Omvec)$ is uncollided radiance from (\ref{e:uncollided}) with $z = \tau/\sigma$.

With this formal solution of the integro-differential RTE in hand, we can substitute the definition of $S(\tau,\Omvec)$ in terms of $I(\tau,\Omvec)$ expressed in (\ref{e:nD_RTE_SourceFunction}), and obtain an integral form of the RTE:
\begin{equation}
I(\tau,\Omvec) = \int\limits_{\Xi_d}\int\limits_0^{\tau^\star} \mathcal{K}(\tau,\Omvec;\tau^\prime,\Omvec^\prime) I(\tau^\prime,\Omvec^\prime) \dif\tau^\prime\dif\Omvec^\prime
+ Q_I(\tau,\Omvec),
\label{e:nD_integral_RTE}
\end{equation}
where
\begin{equation}
Q_I(\tau,\Omvec) = \exp(-\tau/\mu_0) \delta(\Omvec-\Omvec_0).
\label{e:Q_integral_RTE}
\end{equation}
This is simply the uncollided radiance field $I_0(\tau,\Omvec)$ from (\ref{e:simpleS_sweep}) and (\ref{e:uncollided}) where, without loss of generality, we henceforth take $F_0 = 1$.  The kernel of the integral RTE is given by
\begin{equation}
\mathcal{K}(\tau,\Omvec;\tau^\prime,\Omvec^\prime) =
\omega p_g(\Omvec\cdot\Omvec^\prime)
\Theta\left(\frac{\tau-\tau^\prime}{\mu}\right)
\frac{\exp(-|\tau-\tau^\prime|/|\mu|)}{|\mu|},
\label{e:K_integral_RTE}
\end{equation}
where $\Theta(x)$ is the Heaviside step function ($=1$ if $x \ge 0$, $=0$ otherwise).  It enforces the causal requirement of doing upwind sweeps.

Conversely, one can substitute (\ref{e:simpleS_sweep}) into (\ref{e:nD_RTE_SourceFunction}), with the adopted change of spatial coordinate ($z\mapsto\tau$) leading to $\sigma_\text{s}\mapsto\omega$.  That yields the so-called ``ancillary'' integral RTE:
\begin{equation}
S(\tau,\Omvec) = \int\limits_{\Xi_d}\int\limits_0^{\tau^\star} \mathcal{K}(\tau,\Omvec;\tau^\prime,\Omvec^\prime) S(\tau^\prime,\Omvec^\prime) \dif\tau^\prime\dif\Omvec^\prime
+ Q_S(\tau,\Omvec),
\label{e:nD_ancillary_IRTE}
\end{equation}
where
\begin{equation}
Q_S(\tau,\Omvec) = \omega p_g(\Omvec\cdot\Omvec_0) \exp(-\tau/\mu_0).
\label{e:Q_ancillary_IRTE}
\end{equation}
The kernel is the same as given in (\ref{e:K_integral_RTE}).  However, if there were spatial variations in the optical properties, SSA $\omega$ and/or PF $p(\cdot)$, then the kernels would differ in that (\ref{e:Q_integral_RTE}) would use the starting point and (\ref{e:nD_ancillary_IRTE}) the end point of the transition; see, e.g., \cite{DavisKnyazikhin2005}.

If (\ref{e:nD_integral_RTE}) is written in operator language as $I = K I + Q_I$, then it is easy to verify that the Neumann series is a constructive approach for the solution: $I = \sum_{n=0}^\infty I_n$, where $I_{n+1} = K I_n$, hence
\begin{equation}
I = \sum_{n=0}^\infty K^n Q_I = (E-K)^{-1} Q_I,
\label{e:IRTE_soln}
\end{equation}
where $E$ is the identity operator.  This applies equally to the estimation of $S$ as a solution of (\ref{e:nD_ancillary_IRTE}).  Once $S(\tau,\Omvec)$ is a known quantity, one can obtain the readily observable quantity $I(\tau,\Omvec)$ using (\ref{e:simpleS_sweep}).

\noindent
{\it Comment on Angular Reciprocity:}

\noindent
Note that $\mathcal{K}(\tau,\Omvec;\tau^\prime,\Omvec^\prime)$ in (\ref{e:K_integral_RTE}) is invariant when we replace $(\tau,\Omvec;\tau^\prime,\Omvec^\prime)$ with $(\tau^\prime,-\Omvec^\prime;\tau,-\Omvec)$, i.e., swap positions in the medium and switch the direction of propagation.  This leads to reciprocity of the radiance fields for plane-parallel slab media under the exchange of sources and detectors \cite{Chandra1950}.  In our case, we consider radiance escaping the medium in reflection ($\tau = 0$) or transmission ($\tau = \tau^\star$) since the source is external.  Focusing on reflected radiance in BRF form (\ref{e:BRF_form}), reciprocity reads as
\begin{equation}
I_\text{BRF}(0,\Omvec;\Omvec_0) = I_\text{BRF}(0,-\Omvec_0;-\Omvec),
\label{e:R_reciprocity}
\end{equation}
where the second angular argument reads as a parameter (from upper BC) rather than an independent variable.
Similarly, we have $I_\text{BRF}(\tau^\star,\Omvec;\Omvec_0) = I_\text{BRF}(\tau^\star,-\Omvec_0;-\Omvec)$ in transmittance, using the same BRF-type normalization.

We can verify transmissive reciprocity explicitly on $I_0 = Q_I$ in (\ref{e:Q_integral_RTE}) for uncollided radiance. Reflective reciprocity can be verified less trivially using singly-scattered radiance $I_1 = K I_0 = K Q_I$.  Based on (\ref{e:Q_integral_RTE})--(\ref{e:K_integral_RTE}), this leads to
\begin{equation}
I_1(0,\Omvec;\Omvec_0) =
\omega p_g(\Omvec\cdot\Omvec_0)
\int\limits_0^{\tau^\star} \exp(-\tau^\prime/\mu_0)
\exp(-\tau^\prime/|\mu|) \dif\tau^\prime/|\mu|.
\label{e:1_scatter_sweep}
\end{equation}
From there, (\ref{e:BRF_form}) yields
\begin{equation}
\frac{c_d}{\mu_0} I_1(0,\Omvec;\Omvec_0) = c_d
\frac{\omega p_g(\Omvec\cdot\Omvec_0)}{\mu_0+|\mu|}
\left( 1-\exp\left[-\tau^\star\left(\frac{1}{\mu_0}+\frac{1}{|\mu|}\right)\right] \right),
\label{e:1_scatter_R}
\end{equation}
with $\mu_0 > 0$ and $\mu < 0$.  Noting that $-\mu > 0$ and $-\mu_0 < 0$, (\ref{e:1_scatter_R}) verifies (\ref{e:R_reciprocity}).  The same can be shown for transmitted radiance.

\section{Generalized Radiative Transport in $d$ Spatial Dimensions}
\label{s:dDim_gRTE}

\subsection{Emergence of Non-Exponential Transmission Laws in the Cloudy Atmosphere}
\label{s:nonExpTran}

\subsubsection{Two-Point Correlations in Clouds According to In-Situ Probes}

We refer to Davis and Marshak \cite{DavisMarshak04} and Davis \cite{Davis2006} for a detailed accounts of the optical variability we expect---and indeed observe \cite[and references therein]{Davis_etal99}---in the Earth's turbulent cloudy atmosphere.  See also Kostinski \cite{Kostinski2001} for an interestingly different approach.

The important---almost defining---characteristic of this variability is that it prevails over a broad range of scales, which translates statistically into auto-correlation properties with long ``memories.''  The traditional metric for 2-point correlations in turbulent media is the $q^\text{th}$-order structure function \cite{MoninYaglom1975}
\begin{equation}
\text{SF}_q(r) = \overline{|f(\vect{x}+\vect{r})-f(\vect{x})|^q},
\label{e:qSF_def}
\end{equation}
where $f(\vect{x})$ is a spatial variable of interest, $\vect{r}$ is a spatial increment of magnitude $r$, and the overscore denotes spatial or ensemble averaging.  Structure functions are the appropriate quantities to use for fields that are non-stationary but have stationary increments.\footnote{
Following many others, we borrow here the terminology of time-series analysis since the proper language of statistical ``homogeneity'' might be confused with structural homogeneity, a usage we've already introduced in the above.}  
Stationarity of the increments in $f(\vect{x})$ means that the ensemble average on the right-hand side of (\ref{e:qSF_def}) depends only on $\vect{r}$.  Further assuming statistical isotropy, for simplicity, it will depend only on $r$.  The norm of wavelet coefficients have become popular alternatives to the absolute increment in $f(\vect{x})$ used in (\ref{e:qSF_def}) \cite{Farge92,MuzyEtAl1994}.

As expected for all turbulent phenomena, in-situ observations in clouds invariably show that \cite{Davis_etal94,Davis_etal96,Marshak_etal97,Davis_etal99}
\begin{equation}
\overline{|f(\vect{x}+\vect{r})-f(\vect{x})|^q} \sim r^{\zeta_q},
\label{e:qSF_scaling}
\end{equation}
for $r$ ranging from meters to kilometers, where $\zeta_q$ is generally a ``multiscaling'' or ``multifractal'' property, meaning that $\zeta_q/q$ is not a constant.  Physically, this means that knowledge of one statistical moment, such as variance SF$_2(r)$, of the absolute increments cannot be used to predict all others based on dimensional analysis.  Otherwise, it is deemed ``monoscaling'' or ``monofractal.''

It has long been known theoretically---and well-verified empirically---that $\zeta_2 = 2/3$ when $f$ is a component of the wind \cite{Kolmogorov41}, temperature or a passive scalar density \cite{Obukhov1949,Corrsin1951}, when the turbulence is statistically homogeneous and isotropic.  This is equivalent \cite{MoninYaglom1975} to stating that energy spectra of these various quantities in turbulence are power-law with an exponent $\beta = -5/3$ in (\ref{e:PowerLaw_EnergySpectum}).  It can also be shown theoretically that $\zeta_q$ is necessarily a convex function, a prediction that has also been amply verified empirically, although in practice the convexity is relatively weak.

At scales smaller than meters, cloud liquid water content (LWC) undergoes, according to reliable in situ measurements in marine stratocumulus, an interesting transition toward higher levels of variability than expected from the scaling in (\ref{e:qSF_scaling}) \cite{Davis_etal99}.  Specifically, sharp quasi-discontinuities associated with positively-skewed deviations occur at random points/times in the transect through the LWC field sampled by airborne instruments.  These jumps are believed to be a manifestation of the random entrainment of non-cloudy air into the cloud \cite{GerberEtAl01}.

At sufficiently large scales, $\overline{|f(\vect{x}+\vect{r})-f(\vect{x})|^q}$ ceases to increase with $r$ as $f(\vect{x})$ become independent (decorrelates) from itself at very large distances.  For non-negative properties, such as the extinction coefficient or particle density, this decoupling has to happen at least at the scale $r$ where the absolute increments (fluctuations) become commensurate in magnitude with the positive mean of the property.  This rationalizes the upper limit of the scaling range for cloud LWC or droplet density at the scale of several kilometers.  In the cloudy atmosphere, decorrelation can happen sooner in the vertical than in the horizontal in cases of strong stratification, i.e., stratus and strato-cumulus scenarios versus broken cumulus generated by vigorous convection.

In atmospheric RT applications to be discussed next, the outer limit of $r$ only needs to be on the order of whatever scale it takes to reach significant optical distances.  That can be less than cloud thickness in stratus/strato-cumulus cases, or can stretch to the whole troposphere (cloudy part of the atmosphere) when convection makes the dynamics more 3D than 2D.

\subsubsection{Statistical Ramifications for Cloud Optical Properties}

Our present goal is to quantify the impact of unresolved random spatial fluctuations of $\sigma(\vect{x})$ on macroscopic transport properties such as large-scale boundary fluxes or remotely observable radiances, spatially averaged in the instrument's field-of-view.  In view of the importance of sweep operations in $d$-dimensional RT, we also need to understand the statistics of integrals of $\sigma(\vect{x})$ over a range of distances $s$ in an arbitrary direction $\Omvec$.  This is the optical path along a straight line between points $\vect{x}$ and $\vect{x}+s\Omvec$:
\begin{equation}
\tau(\vect{x},\vect{x}+s\Omvec) = \int\limits_0^s \sigma(\vect{x}+s\Omvec)\dif s.
\label{e:optical_distance}
\end{equation}
Better still, we need to characterize statistically the direct transmission factor $\exp[-\tau(\vect{x},\vect{x}+s\Omvec)]$ that is used systematically in the upwind sweep operation.  Assuming stationarity and isotropy, we define
\begin{equation}
\overline{T}(s) = \overline{\exp[-\tau(\vect{x},\vect{x}+s\Omvec)]} = \overline{\exp[-\sigma_\text{avr}(\vect{x},\Omvec;s)s]},
\label{e:mean_Tdir}
\end{equation}
where $\sigma_\text{avr}(\vect{x},\Omvec;s)$ is the average extinction encountered by radiation propagating uncollided between $\vect{x}$ and $\vect{x}+s\Omvec$:
\begin{equation}
\sigma_\text{avr}(\vect{x},\Omvec;s) = \frac{1}{s}\int\limits_0^s\sigma(\vect{x}+s^\prime\Omvec)\dif s^\prime
\label{e:1ptScaleInv}
\end{equation}
This is essentially a coarse version of the random field $\sigma(\vect{x})$, smoothed over a given scale $s$.  What behavior do we expect it to have?

Being at the core a material density (times a collision cross-section), increments of $\sigma(\vect{x})$ will follow (\ref{e:qSF_scaling}) when $s$ is varied, but the segment-mean $\sigma_\text{avr}(\vect{x},\Omvec;s)$ will not depend much on the scale $s$.  Indeed, comparing values of $\sigma_\text{avr}(\vect{x},\Omvec;s)$ for different values of $s$ is really just saying, with linear transects, that the notion of a material density can be defined.  That simple proposition is usually stated as a volumetric statement: the amount of material in a volume $s^d$ is proportional to that volume, and the proportionality factor is called ``density.''

Even more fundamentally, $s$-independence of (\ref{e:1ptScaleInv}), at least in the $s \to 0$ limit,\footnote{
This limit is to be understood physically as to the scale where noise-like fluctuations occur, which is at least the inter-particle distance in a cloud but could be larger \cite{Davis_etal99,GerberEtAl01}.}
is tantamount to saying that $\sigma(\vect{x}) = \sigma_\text{avr}(\vect{x},\Omvec;0)$ is indeed a ``function,'' i.e., the symbol ``$\sigma(\vect{x})$'' represents a number.  This is a natural consequence of increments that vanish at least on average with $s$, a property known as ``stochastic continuity,'' which incidentally does not exclude a countable number of discontinuities (e.g., sharp cloud edges in the atmosphere).
All of these ramifications come with the above-mentioned long-range correlations described by (\ref{e:qSF_scaling}) for finite positive values of $\zeta_q$.

A counter-example of such intuitive $s$-independent behavior for averages is the spatial equivalent of white noise.  Indeed, if $\sigma(\vect{x})$ on the right-hand side of (\ref{e:1ptScaleInv}) could somehow represent white noise (in the discrete world, just a sequence of uncorrelated random numbers), then the left-hand side is just an estimate of its mean value based on as many samples as there are between 0 and $s$.  As $s\to\infty$, this estimate is known to converge to the mean (law of large numbers) in $1/\sqrt{s}$ and, moreover, the PDF of $\sigma_\text{avr}(\vect{x},\Omvec;s)$ is a Gaussian with variance $\propto s$ (central limit theorem).  This is a reminder that one should not denote white (or other) noises as a function $f(\vect{x})$ but rather as a distribution that exists only under integrals: $f(\vect{x})\dif\vect{x}$ is better, $f(\vect{x},\dif\vect{x})$ is the best.  That remark is key to Davis and Mineev-Weinstein's \cite{DavisMineev11} generalization of RT to rapidly fluctuating extinction fields, possibly with anti-correlations across scales.

\subsubsection{Representation of Spatial Fluctuations with Gamma PDFs}

Turbulent density fields are often found to have log-normal distributions and cloud LWC is no exception.  However, Gamma distributions can be used to approximate log-normals in terms of skewness, and are much easier to manipulate.  Barker et al. \cite{Barker1996b} showed that Gamma distributions with a broad range of parameters can fit histograms of cloud optical depth reasonably well.  Now, letting $\vect{x} = (\vec{x},z)^\text{T}$, cloud optical depth is simply $\sigma(\vec{x},z)$ averaged spatially over $z$ from 0 to the fixed cloud (layer) thickness $H$, then re-multiplied by $H$, and histograms were cumulated over $\vec{x}$; in this case, many large cloudy images where $\vec{x}$ designates a 30-m LandSat pixel.  Barker \cite{Barker1996a} proceeded to apply this empirical finding to evaluate unresolved variability effects using the standard two-stream approximation for scattering media, as defined further on, in (\ref{e:2st_model}), for a representative case.  We have used it previously for uncollided radiation \cite{Davis2006}, and do the same here.

In summary, assume that the random number $\sigma_\text{avr}(\vect{x},\Omvec;s)$ in (\ref{e:mean_Tdir})--(\ref{e:1ptScaleInv}) is indeed statistically independent of $s$, at least over a range of values that matter for RT.  This range of $s$ should encompass small, medium and large propagation distances between emission, scattering, absorption or escape events.  These events can unfold in the whole medium, or else in portions of it we might wish to think of in isolation, e.g., clouds in the atmosphere.  If $\overline{\sigma}$ is the spatially-averaged extinction coefficient, over some or all of the transport space $(\vect{x},\Omvec)$, then we require the range of statistical independence on $s$ to go from vanishingly small to several times $1/\overline{\sigma}$.   Were the medium homogeneous, this last quantity would be the particle's MFP, but is in general an underestimate of the true MFP \cite{DavisMarshak04}, as illustrated further on in the specific case of interest here.

Following Barker et al. \cite{Barker1996b}, we now assume that the variability of $\sigma_\text{avr}(\vect{x},\Omvec;s)$, for fixed $s$, can be approximated with an ensemble of Gamma-distributed values:
\begin{equation}
\Pr\{\sigma,\dif\sigma\} = \frac{(a/\overline{\sigma})^a}{\Gamma(a)} \, \sigma^{a-1}\exp(-a\sigma/\overline{\sigma}) \, \dif\sigma,
\label{e:Gamma_pdf}
\end{equation}
where $\overline{\sigma}$ is the mean and $a$ is the variability parameter
\begin{equation*}
a = \frac{1}{\overline{\sigma^2}/\overline{\sigma}^2-1},
\end{equation*}
an important quantity that varies from 0$^+$ to $\infty$ since $\overline{\sigma^2} \ge \overline{\sigma}^2$ (Schwartz's inequality).  If $\sigma_\text{avr}(\vect{x},\Omvec;s)$ is $\Gamma$-distributed for fixed $s$, then so is their product $\tau(\vect{x},\vect{x}+s\Omvec)$ in (\ref{e:optical_distance}).

Equation (\ref{e:mean_Tdir}) then reads as the Laplace characteristic function of this Gamma PDF supported by the positive real axis:
\begin{equation}
T_a(s) = \int\limits_0^\infty\exp(-\sigma s)\Pr\{\sigma,\dif\sigma\} = \frac{1}{(1+\overline{\sigma}s/a)^a}.
\label{e:Gamma_Tdir}
\end{equation}
In the limit $a\to\infty$, variance $\overline{\sigma^2}-\overline{\sigma}^2$ vanishes as the PDF in (\ref{e:Gamma_pdf}) becomes degenerate, i.e., $\delta(\sigma-\overline{\sigma})$.  We then retrieve Beer's law: $T_\infty(s) = \exp(-\overline{\sigma}s)$. 
For an explicit model-based derivation of (\ref{e:Gamma_Tdir}) where the exponent $a$ is expressed in terms of the statistical parameters, see Davis and Mineev-Weinstein's \cite{DavisMineev11} study of scale-invariant media in the ``red-noise'' limit ($\beta \to 1$) where the correlations are at their longest.

\begin{figure}[ht]
\begin{center}
\includegraphics[width=5.5in]{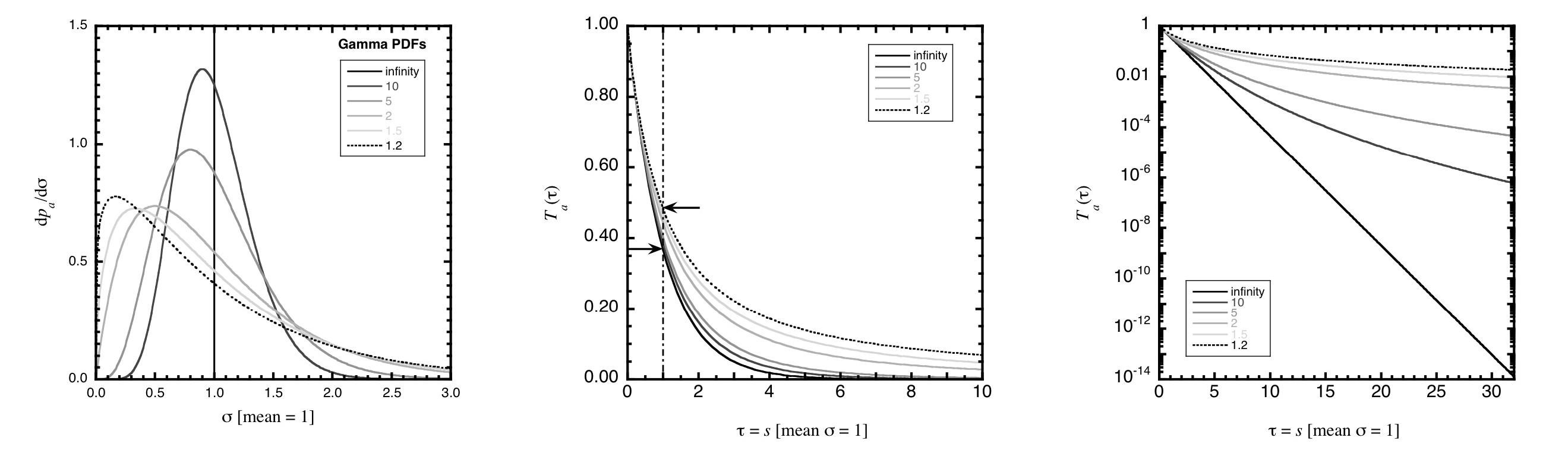}
\end{center}
\caption[new_Tdir]
{\label{f:new_Tdir}
Left to right: Gamma PDFs in (\ref{e:Gamma_pdf}) for the spatial variability of $\sigma$ at a fixed scale $s$, normalized to its ensemble mean $\overline{\sigma}$, for indicated values of $a$; generalized transmission laws $T_a(\tau)$ in (\ref{e:Gamma_Tdir}) associated with PDFs with $\tau = \overline{\sigma}s$; same as middle panel but in semi-log axes and a broader range of $\tau$.}
\end{figure}

The direct transmission law in (\ref{e:Gamma_Tdir}) with a power-law tail thus generalizes the standard law of exponential decay for the cumulative probability of radiation to reach a distance $s$ (or \emph{mean} optical distance $\tau(s)$) from a source without suffering a collision in the material.  Figure~\ref{f:new_Tdir} illustrates both the positively skewed PDFs for $\sigma$, at fixed $s$, in (\ref{e:Gamma_pdf}) and the generalized transmission laws in (\ref{e:Gamma_Tdir}) for selected values of $a$ that we will use further on in numerical experiments.  In the middle panel, we can see that direct transmission probability at $\tau = 1$ increases from $1/\me = 0.368\cdots$ to almost 1/2 going from $a = \infty$ (Beer's exponential law) to a power-law with $a = 1.2$.  The rightmost panel shows that there is still appreciable probability of direct transmission when $a < \infty$ at large optical distances where radiation is all but extinguished in the standard $a = \infty$ case.

Now, viewing $s$ as a random variable that is crucial to transport theory, we have
\begin{equation}
T_a(s) = \Pr\{\text{step}>s\} = \int\limits_s^\infty p_a(s)\dif s.
\label{e:Tdir_asPDF}
\end{equation}
The PDF for a random step of length $s$ is therefore
\begin{equation}
p_a(s) = \left|\frac{\dif T_a}{\dif s}\right|(s) = -\,\frac{\dif T_a}{\dif s}(s) = \frac{\overline{\sigma}}{(1+\overline{\sigma}s/a)^{a+1}}.
\label{e:Gamma_stepPDF}
\end{equation}

In the case of particle transport, we know that the MFP for the $a = \infty$ case (uniform optical media) is $\ell_\infty = 1/\overline{\sigma}$.  What is it for finite $a$ (variable optical media)?  One finds
\begin{equation*}
\ell_a = \langle s \rangle_a = \int\limits_0^\infty s \, \dif T_a(s)
        = \int\limits_0^\infty s \, p_a(s)\dif s
        = \frac{a}{a-1} \, \ell_\infty,
\end{equation*}
which is larger than $\ell_\infty$ and indeed diverges as $a \to 1^+$.  Generally speaking, the step moment $\langle s^q \rangle_a$ is convergent only as long as $-1< q < a$.  This immediately opens interesting questions (addressed in depth elsewhere \cite{DavisMarshak1997} and briefly discussed further on) about the diffusion limit of this variability model when $a \le 2$, i.e., when the 2$^\text{nd}$-order moment of the step distribution is divergent.

\begin{figure}[ht]
\begin{center}
\includegraphics[width=5.5in]{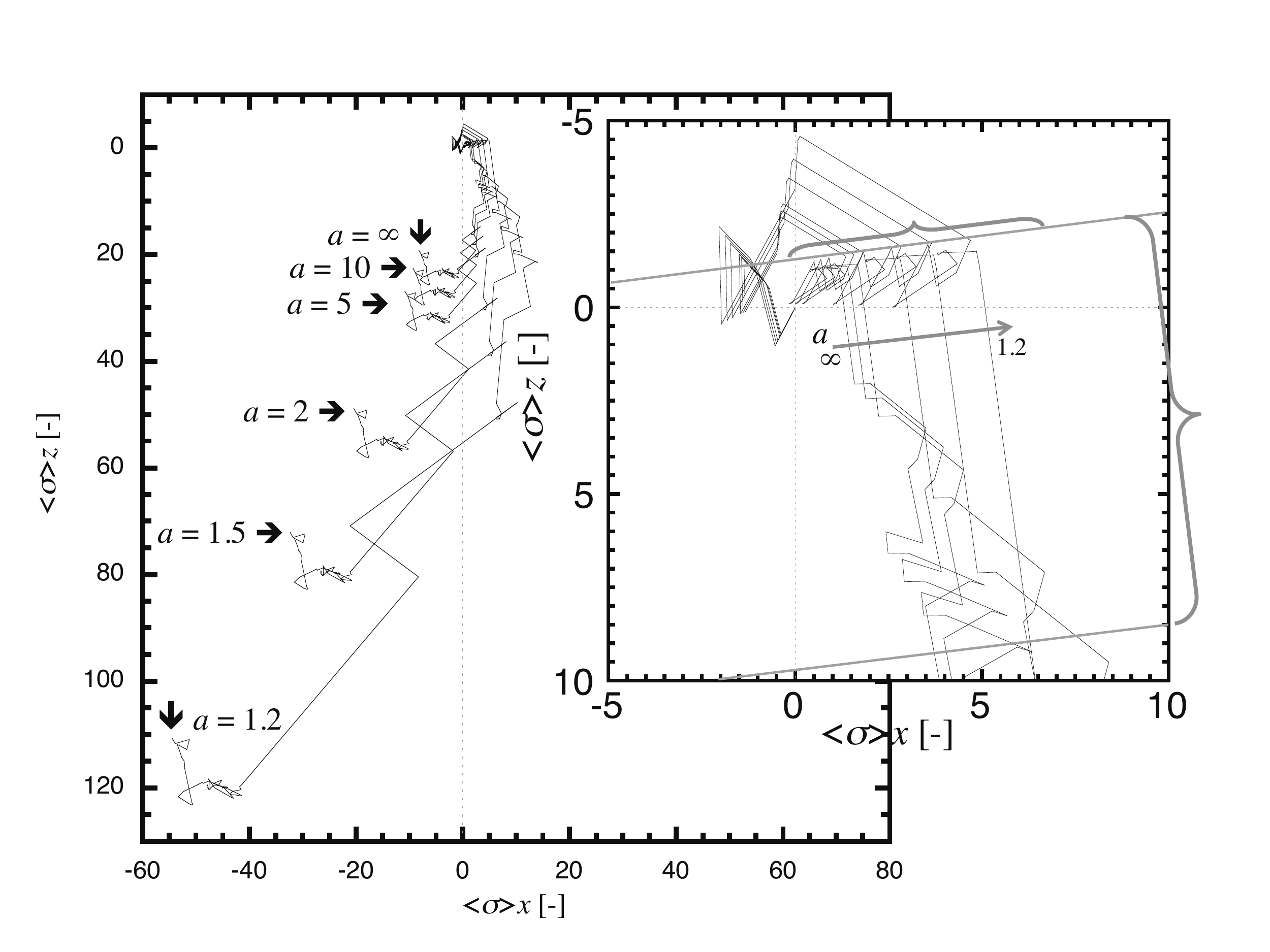}
\end{center}
\caption[RWs_g0_inset]
{\label{f:RWs_g0_inset}
Six traces of random walks in $d=2$ dimensions with 100 isotropic scatterings and step sequences that follow power-law cumulative probabilities (\ref{e:Gamma_Tdir}) and PDFs (\ref{e:Gamma_stepPDF}).  Both scatterings and steps use the same sequences of uniform random variables.  Values of $a$ are $\infty$ (exponential law), 10, 5, 2, 1.5 and 1.2.  The two last ones are asymptotically self-similar L\'evy-stable flights (steps with divergent variance), the three former are asymptotically self-similar Gaussian walks (steps with finite variance), and for $a=2$ it is a transition case (steps with log-divergent variance).  The inset is a $\times 3$ zoom into the commun origin of the 6 traces.  More discussion in main text.}
\end{figure}

Figure~\ref{f:RWs_g0_inset} demonstrates how RT unfolds in $d=2$ inside boundless conservatively scattering media where $\overline{\sigma}$ is unitary.  The media are either uniform or stochastic but spatially-correlated in such a way the ensemble average transmission law is of the power-law form in (\ref{e:Gamma_Tdir}).  We consider media with exponential transmission ($a = \infty$, uniform case) and power-law transmission laws with $a$ = 10, 5, 2, 1.5, and 1.2.  Scattering is assumed to isotropic and we follow the random walk of a transported particle for 100 scatterings.  For illustration, the same scattering angles are used in each of the 6 instances.  For the exponential case, the random free paths are generated using the standard rule: $s = -\log\xi/\overline{\sigma}$, where $\xi$ ia a uniform random variable on the interval (0,1).  For the power-law cases, we use
\begin{equation}
s = a \times (\xi^{-1/a}-1)/\overline{\sigma}
\label{e:Gamma_step}
\end{equation}
As for the random scattering angles, we use the same sequence of 100 values of $2\pi\times\xi$.

In the inset fi Fig.~\ref{f:RWs_g0_inset}, we see that all the traces start at the same point in the same direction.  Physically, we can imagine an electron bound to a crystal surface hoping between holes associated with random defects \cite{LuedtkeLandman1999}.  Certain heterogeneous predator-prey and scavenging problems can also lead to 2D transport processes with a mix of small and large jumps \cite{BuldyrevEtAl2001}.  We can immediately appreciate how the MFP increases as $a \to 1$.  At the same time, the increasing frequency of large jumps enables the cumulative traces to end further and further from their common origin as $a$ decreases from $\infty$ to nearly unity.

One can read Fig.~\ref{f:RWs_g0_inset} with atmospheric optics in mind, albeit with each isotropic scattering representing $\approx$$(1-g)^{-1}$ forward scatterings \cite{Davis2006}.  Recalling that $g$ is in the range 0.75 to 0.85 in various types of clouds and aerosols, this translates to 4 to 7 scatterings before directional memory is lost.  For all values of $a$ there is a wide distribution of lengths of jumps between scatterings.  However, for large values of $a$, the distance covered by a cluster of small steps can equally well be covered with one larger jump.  This behavior is characteristic of solar radiation trapped in a single opaque cloud.  In contrast, for the smallest values of $a$, it is increasingly unlikely that a cluster of smaller steps can rival in scale a single large jump.  This behavior is typical of solar radiation that is alternatively trapped in clouds and bouncing between them.  In other words, we are looking at a 2D version of a typical trace of a multiply scattered beam of sunlight in a 3D field of broken clouds.

\subsection{$d$-Dimensional Generalized RTE in Integral Form}

Our goal is now to formulate the transport equations that describe RT in media when, as in Fig.~\ref{f:RWs_g0_inset}, we transition from an exponential direct transmission law to a power-law counterpart.

Our starting point is the integral form of the $d$-dimensional plane-parallel RTE in (\ref{e:nD_integral_RTE}); alternatively, (\ref{e:nD_ancillary_IRTE}) paired with (\ref{e:simpleS_sweep}).  These formulations are sufficiently general to describe RT and other linear transport processes.  It gets specific to the standard form of RT theory only when we look at the make up of the kernel $\mathcal{K}$ in (\ref{e:K_integral_RTE}), the source terms $Q_I$ in (\ref{e:Q_integral_RTE}) and $Q_S$ in (\ref{e:Q_ancillary_IRTE}).

Therein, we find exponential functions that describe the propagation part of the transport.  Specifically, we identify
\begin{equation}
T_\infty(\tau) = \exp(-\tau)
\label{e:T_Beer_cum}
\end{equation}
in $Q_I$, assuming $\mu_0 = 1$ for the present discussion.  The subscript $\infty$ notation is consistent with our usage in (\ref{e:Gamma_Tdir}) with $\overline{\sigma} = 1$, which is implicit in a non-dimensionalized 1D RT.  This is Beer's classic law of direct transmission, the hallmark of homogeneous optical media where $\tau(s) = \sigma s$ with, in the present setting, $\sigma \equiv \overline{\sigma}$ (i.e., a degenerate probability distribution for $\sigma$).

In $Q_I$, $T_\infty(\tau)$ is therefore at work as the cumulative probability in (\ref{e:Tdir_asPDF}).  In the kernel $\mathcal{K}$, as well as in $Q_S$, we again find exponential functions, but here we interpret them as a PDF:
\begin{equation}
\left|\frac{\dif T_\infty}{\dif \tau}\right| = -T_\infty(|\tau-\tau^\prime|) = \exp(-|\tau-\tau^\prime|),
\label{e:T_Beer_pdf}
\end{equation}
again assuming $\mu = 1$ for the present discussion.  The fact that the (\ref{e:T_Beer_cum}) and (\ref{e:T_Beer_pdf}) are identical functions is of course a defining property of the exponential.

What makes us assign the ``cumulative probability'' interpretation of $\exp(-\tau)$ to its use in $Q_I$, and the ``PDF'' interpretation of $\exp(-|\tau-\tau^\prime|)$ to its use in $Q_S$ and $\mathcal{K}$?  The clue is the foundational transport physics.  In $Q_I = I_0$, the uncollided radiation is simply detected at optical distance $\tau$.  It could have gone deeper into the medium before suffering a scattering, an absorption, a reflection, or escaping through the lower boundary.  In $\mathcal{K}$ however, it is used to obtain $I_n$ from $I_{n-1}$, as previously demonstrated for $n = 1$, cf. (\ref{e:1_scatter_sweep}).  In this case, the radiation must be stopped between $\tau$ and $\tau+\delta\tau$.  It is a probability density that is invariably associated with the differential $\dif\tau$.  Similarly in $Q_S$, the transport process is to stop the the propagation in a given layer and, moreover, it is specifically by a scattering event.

In order to account for unresolved random-but-correlated spatial variability of extinction $\sigma(\vect{x})$, we propose for the integral forms of the $d$-dimensional plane-parallel RTE the following \emph{generalization}: use
\begin{equation}
\mathcal{K}(\tau,\Omvec;\tau^\prime,\Omvec^\prime) =
\omega p_g(\Omvec\cdot\Omvec^\prime)
\Theta\left(\frac{\tau-\tau^\prime}{\mu}\right)
\frac{|\Dot{T}_a(|\tau-\tau^\prime|/|\mu|)|}{|\mu|},
\label{e:K_integral_gRTE}
\end{equation}
rather than (\ref{e:K_integral_RTE}), with
\begin{equation}
Q_S(\tau,\Omvec) = \omega p_g(\Omvec\cdot\Omvec_0) |\Dot{T}_a(\tau/\mu_0)|
\label{e:Q_ancillary_gIRTE}
\end{equation}
for (\ref{e:nD_ancillary_IRTE}), and
\begin{equation}
Q_I(\tau,\Omvec) = T_a(\tau/\mu_0) \delta(\Omvec-\Omvec_0)
\label{e:Q_integral_gRTE}
\end{equation}
for (\ref{e:nD_integral_RTE}), where $a$ can have any strictly positive value, including $\infty$.  We use the overdot notation in (\ref{e:K_integral_gRTE})--(\ref{e:Q_ancillary_gIRTE}) to denote the derivative of a function of a single variable, which is the case here when $\overline{\sigma}$ is combined with $s$ to form $\tau$ in (\ref{e:Gamma_Tdir}) and (\ref{e:Gamma_stepPDF}), and $a$ is viewed as a fixed parameter.

\subsection{Are There Integro-Differential Counterparts of Generalized Integral RTEs?}
\label{s:id_gRTEs}

In short, the $d$-dimensional stochastic transport model we propose is simply to replace $T_\infty(\tau)$ in (\ref{e:T_Beer_cum}) with $T_a(\tau)$ for finite $a$, which we equate with $T_a(s)$ in (\ref{e:Gamma_Tdir}) when $\overline{\sigma}=1$ (thus $s=\tau$).  This logically requires the use of $\dot{T}_a(\tau)$ obtained similarly from $\dif T_a(s)/\dif s$ in (\ref{e:Gamma_stepPDF}).  We thus have a well-defined transport problem using an integral formulation, to be solved analytically or numerically.  Now, is there an integro-differential counterpart?

We do not yet have an answer to this question.  One path forward to address it is to follow the steps of Larsen and Vasques \cite{LarsenVasques11} who start with the classic RT/linear Boltzmann equation in integro-differential form and transform it into a ``non-classical'' one by introducing a special kind of time-dependence that is essentially reset to epoch 0 at every scattering.  Non-exponential free path distributions are thus accommodated, and a modified diffusion limit is derived in cases where $\langle s^2 \rangle$ is greater than $2\langle s \rangle^2$, its value for the exponential distribution, but not too much larger.  Traine and co-authors \cite{TaineEtAl2010} have also proposed a ``generalized'' RTE for large-scale transport through random (porous) media; this model uses an empirical counterpart of our parametric non-exponential transmission law in some parts of the computation, but retains the standard integro-differential form for the final estimation of radiance using the upwind sweep operator in (\ref{e:S_sweep}).

Another path forward is to essentially define new differential (or more likely pseudo-differential) operators as those from which the new integral operator in (\ref{e:K_integral_gRTE}) follows. This amounts to broadening the definition of the Green function,
$G(\tau,\Omvec;\tau_\star,\Omvec_\star) =
T_a(|\tau-\tau_\star|/|\mu|)\delta(\Omvec-\Omvec_\star)$,
for 1D RT in the absence of scattering, previously with $a = \infty$, now with arbitrary values, and assigning a role to $\partial G/\partial\tau$.  This more formal approach seems to us less promising in terms of physical insights---a judgment that may be altered if a rigorous connection to the concept of fractional derivatives \cite{MillerRoss1993} can be established.  These pseudo-differential operators have indeed found many fruitful applications in statistical physics \cite{MetzlerKlafter2000,West2003}.

Although out of scope for the present study, there is an implicit time-dependence aspect to generalized (as well as standard) RT even if the radiance fields are steady in time.  The best way to see this is to return to the inset in Fig.~\ref{f:RWs_g0_inset}.  The highlighted region (between gray brackets) shows in essence how standard and generalized 2D RT unfolds for solar illumination of a medium of optical thickness $\approx$11 at an angle of $\approx$30$^\circ$ from zenith.  The smaller the value of $a$, the shorter the path of the light inside the medium.  The number of scatterings decreases from 25 ($a=\infty$) to 10 ($a=1.2$).  The flight time for sunlight to cross the cloudy portion of the atmosphere---at most from near-ground level to the troposphere (10 to 15~km altitude, depending on latitude)---cannot be measured directly.  However, it can be estimated statistically via oxygen spectroscopy \cite{Pfeil_etal98}.  Pfeilsticker, Scholl et al. \cite{Pfeil99,Scholl_etal06}, as well as Min, Harrison et al. \cite{MinHarrison99,Min_etal01,Min_etal04}, have found that the more variable the atmosphere at a given mean optical thickness, the shorter the top-to-ground paths on average.  This finding offers a degree of validation of generalized RT for applications to the Earth's cloudy atmosphere.

In the remainder of this study, we derive analytical and numerical solutions of the generalized RTE in (\ref{e:nD_integral_RTE}) with (\ref{e:Q_integral_gRTE})--(\ref{e:K_integral_gRTE}), and then apply them to specific topics where standard and generalized RT differ significantly.

\section{Deterministic Numerical Solution in $d = 1$: The Markov Chain Approach}
\label{s:2stream_MarCh}

In \S\ref{s:dDim_sRTE}, we stated that once we adopted the H--G PF in Table~\ref{t:Definitions} the whole 1D RT problem is determined entirely (in the absence of surface reflection) by three numbers, $\{\omega,g;\tau^\star\}$ for a given $d$ = 1, 2, or 3, with the possible addition of $\mu_0$ when $d > 1$.  To this small parameter set, we now add the exponent $a$ of the power-law direct transmission function that distinguishes standard RT (exponential limit, $a\to\infty$) from its generalized counterpart ($0 < a < \infty$).  The complete parameter set is therefore $\{\omega,g,a;\tau^\star(;\mu_0)\}$.

\subsection{Exact Solution of the Standard RTE in $d = 1$}

The ``$d=1$'' (\emph{literal} 1D) version of 1D RT has in fact a vast literature of its own since as it is formally identical to the two-stream RT model \cite{Schuster1905,KubelkaMunk1931}, a classic approximation for (standard) RT in $d=3$ space.  This simplified RT model is still by far the most popular way to compute radiation budgets in climate and atmospheric dynamics models \cite{MeadorWeaver80}.  We note that there is no longer an angular integral to compute in the $d$-dimensional RTE in (\ref{e:1D_RTE}).  It is understood to be replaced everywhere by a sum over two directions: ``up'' and ``down.'' Correspondingly, scattering can only be through an angle of $0$ or $\pi$ rad: $\mu_\text{s} = \pm 1$, respectively.  The $d=1$ RT problem at hand thus takes the form of a pair of coupled ODEs:
\begin{equation}
\left( \pm\frac{\dif\;}{\dif\tau} + 1 \right) I_{\pm}(\tau) =
\omega \left[p_{+}I_{\pm}(\tau)+p_{-}I_{\mp}(\tau)\right] + q_{\pm}(\tau)
\label{e:2st_model}
\end{equation}
with $p_{\pm} = (1 \pm g)/2$ (cf. Table~\ref{t:Definitions}) and $q_{\pm}(\tau) = \omega p_{\pm} \exp(-\tau)$.  This system of coupled ODEs is subject to BCs $I_{+}(0) = I_{-}(\tau^\star) = 0$ when $\rho = 0$ (otherwise $I_{-}(\tau^\star) = \rho I_{+}(\tau^\star)$).

Let us use
\begin{equation}
I_{\pm}(\tau) = \frac{J(\tau) \pm F(\tau)}{2}
\label{e:1D_diffusion}
\end{equation}
to recast the diffuse radiance field in the above 2-stream model, where
\begin{eqnarray}
\label{e:J_2st_def}
J(\tau) &=& I_+(\tau) + I_-(\tau), \\
\label{e:F_2st_def}
F(\tau) &=& I_+(\tau) - I_-(\tau),
\end{eqnarray}
are respectively the \emph{scalar} and \emph{vector} fluxes.

By summing the two ODEs in (\ref{e:2st_model}), we find an expression of radiant energy conservation:
\begin{equation}
\dif F/\dif\tau = -(1-\omega)J + \omega\exp(-\tau).
\label{e:2st_conservation}
\end{equation}
Differencing (\ref{e:2st_model}) yields
\begin{equation}
F(\tau) = (-\dif J/\dif\tau + \omega g \me^{-\tau})/(1-\omega g).
\label{e:2st_model_mFick}
\end{equation}
The 1st term on the right-hand side (and the only one that survives after the 2nd one has decayed at large $\tau$) is a non-dimensional version of Fick's law, a reminder that diffusion theory is exact in $d=1$.  Using (\ref{e:2st_model_mFick}) in (\ref{e:2st_conservation})
leads to a 1D screened Poisson equation for $J(\tau)$:
\begin{equation*}
\left[ -\frac{\dif^2\;}{\dif\tau^2} + (1-\omega)(1-\omega g) \right] J(\tau) =
\omega \left[ 1+(1-\omega)g \right] \exp(-\tau),
\end{equation*}
subject to BCs, $J(0)+F(0) = J(\tau^\star)-F(\tau^\star) = 0$ when $\rho = 0$ (black surface).  Factoring in (\ref{e:2st_model_mFick}), these are always of the 3rd (Robin) type.

When $\omega = 1$ (no absorption), the solution of the above pair of ODEs and BCs is
\begin{eqnarray}
\label{e:J_2st_NoAbs}
J(\tau) &=& 1+R(\tau^\star)\times\left(1-\frac{\tau}{\tau^\star/2}\right)-\exp(-\tau), \\
\label{e:F_2st_NoAbs}
F(\tau) &=& T(\tau^\star)-\exp(-\tau).
\end{eqnarray}
We have used here boundary-escaping radiances
\begin{eqnarray}
\label{e:R_2st_NoAbs}
R(\tau^\star) &=& I_{-}(0) = 1-T(\tau^\star), \\
\label{e:T_2st_NoAbs}
T(\tau^\star) &=& I_{+}(\tau^\star)+\exp(-\tau^\star) = \frac{1}{1+(1-g)\tau^\star/2},
\end{eqnarray}
in the above representation of the solution.  When $\omega < 1$, somewhat more complex expressions result in the form of 2nd-order rational functions of $\exp(-k\tau)$, where $k = 1/\sqrt{(1-\omega)(1-\omega g)}$, with polynomial coefficients dependent on $\omega$, $g$ and $\exp(-\tau)$.  All these classic results will be used momentarily to verify the new Markov chain numerical scheme.

\subsection{Markov Chain (MarCh) Scheme}

We now adapt our ``Markov Chain'' (MarCh) formulation of standard RT in $d=3$ dimensions \cite{Xu_EtAl_MarCh11,Xu_EtAl_MarChSurf11,Xu_EtAl_MarChLin12} to the present $d=1$ setting for generalized RT.  MarCh is an under-exploited alternative to the usual methods of solving the plane-parallel RT problem first proposed by Esposito and House \cite{Esposito_House78,Esposito79}.  It differs strongly from many of the usual approaches: discrete ordinates, spherical harmonics, adding/doubling, matrix-operator and kindred techniques.  It has more in common with source iteration (successive orders of scattering, or Gauss-Seidel iteration), and even with Monte Carlo (MC).  In short, we can say that MarCh is an efficient deterministic solution of a discretized version of the integral RTE solved by MC.  We illustrate in $d=1$ for simplicity, but also for previously articulated reasons, that there may be an acute need for generalized RT in the 2-stream approximation in climate and, generally speaking, atmospheric dynamical modeling.

%
The generalized ancillary integral RTE is expressed in generic form in (\ref{e:nD_ancillary_IRTE}) with the kernel in (\ref{e:K_integral_gRTE}) and the source term in (\ref{e:Q_ancillary_gIRTE}).  In $d = 1$, it yields a system of two coupled integral equations for the two possible directions in $S_\pm(\tau)$:
\begin{eqnarray}
S_\pm(\tau) &=& \omega \left[
p_\pm \int\limits_0^\tau
S_{+}(\tau^\prime) \, \left|\dot{T}_a(\tau-\tau^\prime)\right| \, \dif\tau^\prime +
p_\mp \int\limits_\tau^{\tau^\star}
S_{-}(\tau^\prime) \, \left|\dot{T}_a(\tau^\prime-\tau)\right| \, \dif\tau^\prime
\right] \nonumber \\
&+& Q_{S\pm}(\tau),
\label{e:AncIRTE_1D}
\end{eqnarray}
where 
\begin{equation}
Q_{S\pm}(\tau) = \omega p_\pm \left|\dot{T}_a(\tau)\right|.
\label{e:AncIRTE_1D_Q_Order1}
\end{equation}
We recognize here the operator form of the integral equation, $S = K S + Q_S$, which can be solved by Neumann series expansion, similarly to (\ref{e:IRTE_soln}):
\begin{equation}
S = Q_S + K Q_S + K^2 Q_S + \cdots = (E-K)^{-1}Q_S.
\label{e:Neumann_S}
\end{equation}

As detailed in the Appendix, the pair of simultaneous integral RTEs in (\ref{e:AncIRTE_1D}), given (\ref{e:AncIRTE_1D_Q_Order1}),
are finely discretized in $\tau$ (200 layers with $\Delta\tau = 0.05$, hence $\tau^\star = 10$), with careful attention to accuracy in the evaluation of the integrals using finite summations.  The resulting  matrix problem is large but tractable.  It can be solved using either a truncated series expansion of matrix multiplies or the full matrix inversion depending on problem parameters (primarily, $\tau^\star$) and the desired accuracy.

As usual when working with the ancillary integral RTE, we finish computing radiance \emph{detected} inside the medium using the formal solution, as in (\ref{e:simpleS_sweep}) but in $d = 1$ format, and with the appropriate generalized transmission law:
\begin{equation}
\left\{ \begin{array}{l}
I_{+}(\tau) = \int\limits_0^\tau
S_{+}(\tau^\prime) \, T_a(\tau-\tau^\prime) \, \dif\tau^\prime
+ T_a(\tau),
\\
I_{-}(\tau) = \int\limits_\tau^{\tau^\star}
S_{-}(\tau^\prime) \, T_a(\tau^\prime-\tau) \, \dif\tau^\prime,
\end{array} \right.
\label{e:FormalSoln_1D_escape}
\end{equation}
with $q_{-}$.  Indeed, ``detection'' implies that the radiation reaches a level, but could have gone further.  A special case of detection is radiation \emph{escaping} the medium at a boundary: $I_{+}(\tau^\star)$ or $I_{-}(0)$, which can also be obtained from known values of $S_\pm$ using one or another of the expressions in (\ref{e:FormalSoln_1D_escape}).  At any rate, it is the ``cumulative probability'' version of the transmission law that is needed here.
In short, after implementing (\ref{e:Neumann_S}), the final step of the numerical computation is to derive radiances $I_\pm(\tau)$ everywhere (it is required) from the known source function $S_\pm(\tau)$ using a discretized version of (\ref{e:FormalSoln_1D_escape}).

In the Appendix, the discrete-space version of the above problem is derived directly from an analogy with random walk theory using Markov chain formalism: present state, state transition probabilities, probability of stagnation, of absorption (including escape), starting position/direction of walkers, and so on.  Although intimately related to all these concepts, which are used extensively in MC modeling, the new model is deterministic since it uses normal rather than random quadrature rules.  We naturally call it the Markov Chain (MarCh) approach to RT.  In a recent series of papers \cite{Xu_EtAl_MarCh11,Xu_EtAl_MarChSurf11,Xu_EtAl_MarChLin12}, we have brought it to bear on aerosol remote sensing on Earth (in $d=3$), so far only with $a=\infty$, but including polarization.

\subsection{Illustration with Internal Fields}

\begin{figure}[ht]
\begin{center}
\includegraphics[width=4in]{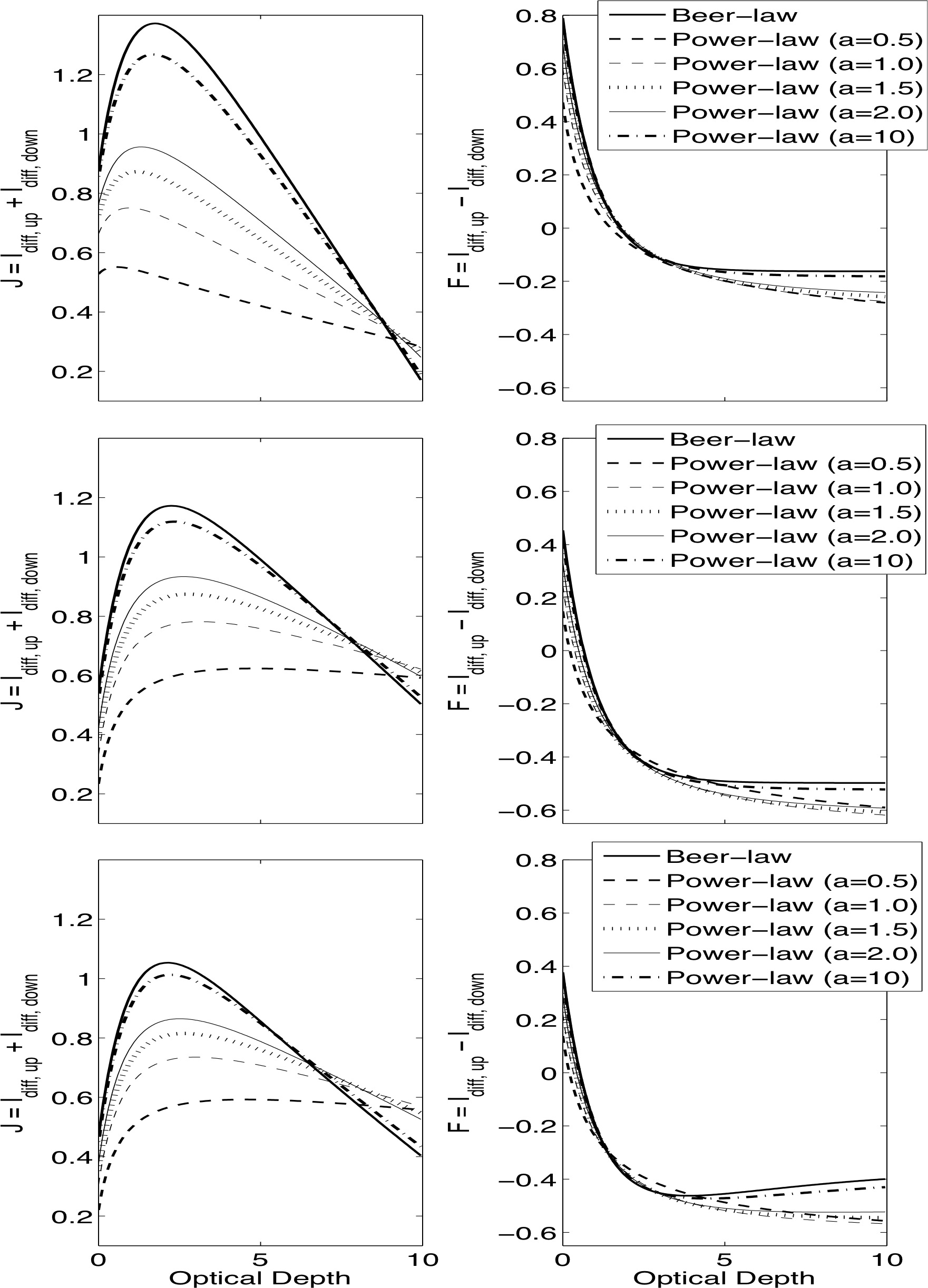}
\end{center}
\caption[MarCh]
{\label{f:MarCh}
Internal radiance fields $J = I_{+}+I_{-}$ (right) and $-F = -I_{+}+I_{-}$ (left) computed using the new MarCh scheme for $d = 1$ described in the Appendix.  $J(\tau)$ in (\ref{e:J_2st_def}) and $-F(\tau)$ in (\ref{e:F_2st_def}) are plotted as a function of optical depth $\tau$ into a medium with $\tau^\star = 10$, the unitary source being at $\tau = 0$, for selected values of $a$.  The standard exponential law obtained when $a\to\infty$ is designated as ``Beer's law.''  In the top two rows, no absorption is included but the phase function is varied: $p_{+} = 1/2$ ($g = 0$) on top; $p_{+} = 0.9$ ($g = 0.8$) in the middle.  In the bottom row, again $g = 0.8$ but $\omega$ is reduced from unity to 0.98.}
\end{figure}

To demonstrate our MarCh code for generalized transport in $d=1$, we focus on uniform or stochastic media with $\tau^\star = 10$ irradiated by a unitary source at its upper ($\tau = 0$) boundary, here, to the left of each panel in Fig.~\ref{f:MarCh}.  We first assume conservative ($\omega = 1$) and isotropic ($g = 0$) scattering.  The outcome is plotted in the top two panels in the $d=1$ equivalent of a decomposition in Fourier modes (in $d=2$) or spherical harmonic modes (in $d=3$).  Specifically, we have scalar flux $J = I_{+}+I_{-}$ in the left column and (negative) vector flux $-F = I_{-}-I_{+}$ in the right column.  In the middle row, $g$ is raised from 0 to 0.8.  In the bottom row, $\omega$ is then lowered from unity to 0.98.  In all of these scenarios, $a$ was varied, the selected values being 1/2, 1, 3/2, 2, 10, and $\infty$; the latter choice is designated as ``Beer's law'' and the others as ``power laws'' in the Figure.

When $\omega = 1$, radiant energy conservation requires that total net flux $F(\tau)+T_a(\tau)$ be constant across the medium, and equal to $T(1,g,a;\tau^\star)$.  This was verified numerically for all values of $a$ and both values of $g$.  In the right hand panels, we see that indeed $-F(\tau) = -T(1,g,a;\tau^\star)+T_a(\tau)$; see (\ref{e:F_2st_NoAbs}) and (\ref{e:T_2st_NoAbs}) for the case of $a = \infty$ and $\omega = 1$.

For reference, Table~\ref{t:MarCh_results} gives $R$, $T_\text{dif}$, $T_a(\tau^\star)$, and absorbtance $A = 1-(R+T) = 1-R-T_\text{dif}-T_a(\tau^\star)$ for our three choices of $\{\omega,g\}$ and all values of $a$.  All the entries in Table~\ref{t:MarCh_results} were verified to all the expressed digits using a custom MC code for generalized transport in $d = 1$.  Apart from the fact that there is no oblique illumination, nor is there a distinction between collimated and diffuse illumination, the key difference between a MC for $d=1$ and $d>1$ is how to select a scattering angle.  In $d>1$, it is a continuous random variable but in $d=1$ the forward versus backward scattering decision is made based on a Bernoulli trial.

As another element of verification for the MarCh code, we recognize in the upper and middle left-hand panels of Fig.~\ref{f:MarCh} the characteristic result for $J(\tau)$ in the case of standard transport theory ($a\to\infty$) in the absence of absorption ($\omega = 1$), namely, the linear decrease modulated by an exponential expressed in (\ref{e:J_2st_NoAbs}).

In standard transport theory in $d = 1$, or using the 2-stream/diffusion approximation for higher dimensions, the linear decrease of $J(\tau)$ when $\omega = 1$ follows directly from the constancy of $F(\tau)$, assuming they include both diffuse and uncollided radiation; see (\ref{e:2st_model_mFick}) and (\ref{e:F_2st_NoAbs}), but without the exponential terms.  An interesting finding here is that, although $F(\tau)+T_a(\tau)$ is constant for all values of $a$, the linear decrease of $J(\tau)+T_a(\tau)$ does not generalize from $a = \infty$ to $a < \infty$.  We conclude that generalized RT conserves energy, as it should, both globally ($A+R+T = 1$) and locally, as expressed in
\begin{equation}
\dif F/\dif\tau = -(1-\omega)J(\tau) + \omega \dot{T}_a(\tau),
\label{e:Gen_2st_model_EnCons}
\end{equation}
which is follows directly from (\ref{e:2st_model})--(\ref{e:F_2st_def}) when $a = \infty$.  It is not obvious how to derive (\ref{e:Gen_2st_model_EnCons}) for the generalized transport model described by one or another of its integral RTEs, even in $d=1$.  In contrast, Fick's law in (\ref{e:2st_model_mFick}), which relates $F$ to $\dif J/\dif\tau$, can only be exact when $a = \infty$ and, moreover, when $d = 1$.

Another interesting numerical finding is that when $\omega = 1$ and $a = \infty$, $T$ depends only on the scaled optical thickness $(1-g)\tau^\star$, as is readily seen in (\ref{e:T_2st_NoAbs}).  This means that, by similarity, an isotropically scattering medium ($g = 0$) with $\tau^\star = 10$ has the same total transmittance $T$ as a forward scattering medium with (say) $g = 0.8$ and $\tau^\star = 50$.  Formally, $T(1,g,\infty;\tau^\star)$ is only a function of $(1-g)\tau^\star$.  More generally, allowing $\omega \le 1$, we have
\begin{equation}
F(\omega,g,\infty;\tau^\star) \equiv
f_F\left(\frac{1-\omega}{1-\omega g},(1-\omega g)\tau^\star\right)
\end{equation}
for $F = A,R,T$, where the first argument on the r.-h. side is known as the similarity parameter \cite{King1987}.  This is not the case when $a < \infty$.

\begin{table}[ht]
\caption{Boundary fluxes $R$, $T_\text{dif}$, $T_\text{dir} = T_a(\tau^\star)$, and absorbtance $A$ for the $d = 1$ stochastic medium with $\tau^\star = 10$ used in Fig.~\ref{f:MarCh}.}
\label{t:MarCh_results}
\begin{center}
\begin{tabular}{|c|c|c||c|c|c|c|c|}
\hline
$\omega$ & $g$ & $a$ & $R$ & $T_\text{dif}$ & $T_\text{dir}$ & $T$ & $A$ \\
\hline\hline
1.00 & 0.0 & $\infty$ & 0.833 & 0.167 & 0.000 & 0.167 & 0.000 \\
\hline
1.00 & 0.0 & 10.      & 0.814 & 0.185 & 0.001 & 0.186 & 0.000 \\
\hline
1.00 & 0.0 & 2.0      & 0.727 & 0.245 & 0.028 & 0.273 & 0.000 \\
\hline
1.00 & 0.0 & 1.5      & 0.693 & 0.260 & 0.047 & 0.307 & 0.000 \\
\hline
1.00 & 0.0 & 1.0      & 0.632 & 0.277 & 0.091 & 0.368 & 0.000 \\
\hline
1.00 & 0.0 & 0.5      & 0.500 & 0.282 & 0.218 & 0.500 & 0.000 \\
\hline\hline
1.00 & 0.8 & $\infty$ & 0.500 & 0.500 & 0.000 & 0.500 & 0.000 \\
\hline
1.00 & 0.8 & 10.      & 0.475 & 0.524 & 0.001 & 0.525 & 0.000 \\
\hline
1.00 & 0.8 & 2.0      & 0.378 & 0.594 & 0.028 & 0.622 & 0.000 \\
\hline
1.00 & 0.8 & 1.5      & 0.345 & 0.608 & 0.047 & 0.655 & 0.000 \\
\hline
1.00 & 0.8 & 1.0      & 0.290 & 0.619 & 0.091 & 0.710 & 0.000 \\
\hline
1.00 & 0.8 & 0.5      & 0.192 & 0.590 & 0.218 & 0.808 & 0.000 \\
\hline\hline
0.98 & 0.8 & $\infty$ & 0.422 & 0.401 & 0.000 & 0.401 & 0.176 \\
\hline
0.98 & 0.8 & 10.      & 0.406 & 0.432 & 0.001 & 0.433 & 0.161 \\
\hline
0.98 & 0.8 & 2.0      & 0.335 & 0.524 & 0.028 & 0.552 & 0.112 \\
\hline
0.98 & 0.8 & 1.5      & 0.309 & 0.546 & 0.047 & 0.593 & 0.098 \\
\hline
0.98 & 0.8 & 1.0      & 0.264 & 0.568 & 0.091 & 0.659 & 0.077 \\
\hline
0.98 & 0.8 & 0.5      & 0.179 & 0.557 & 0.218 & 0.775 & 0.046 \\
\hline
\end{tabular}
\end{center}
\end{table}

\section{Diffusion Study in $d = 2$: Theory and Monte Carlo Simulation}
\label{s:DiffusionLimits}

\subsection{Theoretical Predictions}

In this section, we focus on $d$ = 2 spatial dimensions, partly for simplicity (fidelity with Fig.~\ref{f:RWs_g0_inset}, where nothing is happening outside of the depicted $(x,z)^\text{T}$-plane), partly because there are previously-mentioned two-dimensional transport processes on real substrates (including random ones where a stochastic model is in order).  We focus specifically on non-absorbing media ($\omega$ = 1) over an absorbing lower boundary ($\rho$ = 0).  Moreover, we will assume an isotropic source at the upper boundary, that is, BC in (\ref{e:nD_RTE_upperBC_diffuse}) with $F_0$ = 1.

We will investigate transmitted fluxes, both direct and diffuse, their total $T(g,a;\tau^\star)$ being defined in (\ref{e:nD_RTE_upperBC_diffuse}), but ignoring $\mu_0$.  We start with a review of the standard $a=\infty$ case.

In \S\ref{s:dDim_sRTE}, the exact expression for $T(g,\infty;\tau^\star)$ is given in (\ref{e:T_2st_NoAbs}) for $d=1$ where the diffusion ODE model is mathematically exact.  In $d>1$, diffusion is only a physically reasonable approximation to plane-parallel RT for very opaque highly scattering media.  In lieu of (\ref{e:1D_diffusion}), it is based on the 1st-order truncation \begin{equation}
I(\tau,\Omvec) \approx \frac{J(\tau) + d \times F_z(\tau) \mu}{\Xi_d},
\end{equation}
and, in lieu of the first entry in the next-to-last row of Table~\ref{t:Definitions}, we take
\begin{equation}
p_g(\mu_\text{s}) \approx \frac{1+d \times g \mu_\text{s}}{\Xi_d}.
\end{equation}
This leads to Laplace/Helmholtz or Poisson ODEs for $J(\tau)$, respectively for boundary and volume expressions for the sources.  In plane-parallel slab geometry, the BCs are again Robin-type.  When homogeneous (hence sources in the volume), they are conventionally expressed as
\begin{equation*}
\left[ J - \frac{\chi_d}{(1-\omega g)} \frac{\dif J}{\dif\tau} \right]_{\tau = 0} = 0, \;
\left[ J + \frac{\chi_d}{(1-\omega g)} \frac{\dif J}{\dif\tau} \right]_{\tau = \tau^\star} = 0,
\label{e:dD_diffusion_BCs}
\end{equation*}
where $\chi_d$ is the extrapolation length, i.e., boundary values of $J/|\dif J/\dif z|$, expressed in transport MFPs, that is,
\begin{equation}
\ell_\text{t} = 1/(1-\omega g)\sigma.
\end{equation}
Classic values for $\chi_d$ are listed in Table~\ref{t:Definitions} (last row).  In the absence of absorption and using boundary sources, total transmission is
\begin{equation}
T(g,\infty;\tau^\star) \approx \frac{1}{1+\tau_\text{t}^\star/2\chi_d},
\label{e:dD_diffusion_Ttot}
\end{equation}
where $\tau_\text{t}^\star = (1-g)\tau^\star = H/\ell_\text{t}$ is the scaled optical thickness.  This expression is identical to (\ref{e:T_2st_NoAbs}) for $d=1$ ($\chi_1 = 1$), but here we use $\chi_2 = \pi/4$.

Diffusion theory for $a < \infty$ cases is in a far worse state since we do not know yet how to formulate generalized RT in integro-differential form.  What is known is the asymptotic scaling of $T(g,a;\tau^\star)$ with respect to $\tau_\text{t}^\star$.  Based on the appropriate truncation of the Sparre-Anderson law of first returns \cite{Sparre-Anderson1953}, Davis and Marshak \cite{DavisMarshak1997} showed that
\begin{equation}
T(g,a;\tau^\star) \propto {\tau_\text{t}^\star}^{-\alpha/2},
\label{e:scaling_diffusion_Ttot}
\end{equation}
where $\alpha = \min\{2,a\}$ is the L\'evy index.  Recall that $a$ is the generally non-integer value of the lowest order moment of $\langle s^q \rangle$ that is divergent for the power-law step distribution in (\ref{e:Gamma_Tdir}).  Then one of two outcomes occurs:
\begin{itemize}
\item
If $a \ge 2$, hence $\alpha = 2$, then the position of the random walk in Fig.~\ref{f:RWs_g0_inset} is Gaussian (central limit theorem), and standard diffusion theory applies.  As can be seen from (\ref{e:dD_diffusion_Ttot}), the scaling exponent in (\ref{e:scaling_diffusion_Ttot}) is indeed (negative) $\alpha/2 = 1$.
\item
If $a < 2$, hence $\alpha = a$, then the position of the random walk in Fig.~\ref{f:RWs_g0_inset} is L\'evy-stable (generalized central limit theorems), and the diffusion process is ``anomalous.''
\end{itemize}
The predicted scaling in (\ref{e:scaling_diffusion_Ttot}) will occur for any spatial dimensionality.

\subsection{Numerical Results}

Extensive numerical computations were performed in $d=2$ spatial dimensions using a straightforward Monte Carlo scheme.  The goal was to estimate $T(g,a;\tau^\star)$ for a wide range of $\tau$ (0.125 to 4096), two choices for $g$ (0 and 0.85), and a representative selection of values for $a$: 1.2, 1.5, 2, 5, 10, and $\infty$.  We used (\ref{e:Gamma_step}) to sample the distance to the next collision.

The key idiosyncrasies of Monte Carlo simulation of RT in $d=2$ are for the two procedures for generating random angles:
\begin{itemize}
\item
At the departure point of the trajectory, an isotropic source in the angular half-space ($|\theta| < \pi/2$) uses $\sin\theta_0 = 1-2\xi$ (where $\xi$ is a uniform random variable on [0,1]) and $\cos\theta_0 = \sqrt{1-\sin^2\theta_0}$.
\item
If $g \ne 0$, directional correlation is implemented by computing $\theta_{n+1} = \theta_n + \theta_\text{s}$ where $\theta_\text{s} = 2\tan^{-1}\left[ \tan[(\xi-1/2)\pi]\times(1-g)/(1+g) \right]$
based on the corresponding H--G PF from Table~\ref{t:Definitions} for $d = 2$.
\end{itemize}
The remaining operations (boundary-crossing detection and tallies) are similar in $d$ = 1,2,3.

\begin{figure}[ht]
\begin{center}
\includegraphics[width=5.33in]{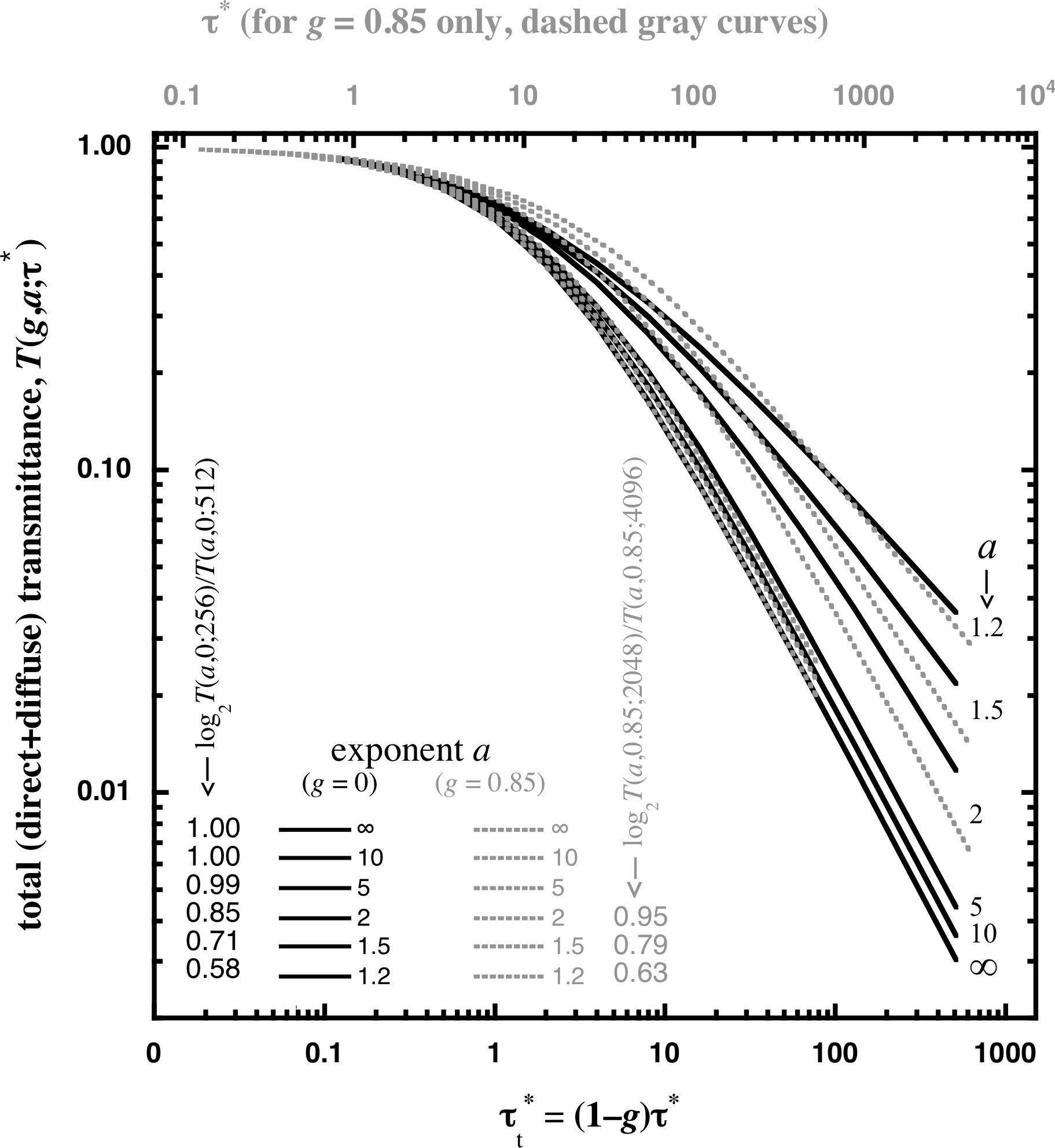}
\end{center}
\caption[2D_T_vs_tau_t]
{\label{f:2D_T_vs_tau_t}
2D Monte Carlo evaluations of $T(g,a;\tau^\star)$ versus transport (or ``scaled'') optical thickness $\tau^\star_\text{t}$ in log-log axes for $g$ = 0 (solid black) 0.85 (dotted gray) and for $a$ = 1.2, 1.5, 2, 5, 10, and $\infty$ (from top down).  Asymptotic scaling exponents are estimated numerically using the last two values of $\tau^\star$ and compared with theoretical predictions in the main text.}
\end{figure}

Figure~\ref{f:2D_T_vs_tau_t} shows our results for $T(g,a;\tau^\star)$ as a function of scaled optical thickness $\tau_\text{t}^\star = (1-g)\tau^\star$ in a log-log plot.  We notice the similarity of $T(0,\infty;\tau^\star)$ and $T(0.85,\infty;\tau^\star)$ using the scaled optical thickness, as predicted in (\ref{e:dD_diffusion_Ttot}): $T(g,\infty;\tau^\star) \sim T((1-g)\tau^\star)$ when $(1-g)\tau^\star \gg 1$.  Specifically, the two transmission curves overlap when plotted against $(1-g)\tau^\star$, at least for large values.  In contrast, we see clear numerical evidence that generalized RT does not have such asymptotic similarity in $T(g,a;\tau^\star)$, as was previously anticipated when examining internal radiation fields in $d=1$.  More precisely, the scaling exponent in (\ref{e:scaling_diffusion_Ttot}) is, as indicated, independent of $g$ but the prefactor (and approach to the asymptote) is.  In Fig.~\ref{f:2D_T_vs_tau_t}, we have estimated the exponents numerically, and they are close to the predicted value, $\min\{1,a/2\}$.

In summary, our modest diffusion theoretical result in (\ref{e:scaling_diffusion_Ttot}) for generalized RT is well verified numerically, and we have gained some guidance about what to expect for a more comprehensive theory.

\section{Single Scattering in $d = 3$: Violation of Angular Reciprocity}
\label{s:ReciprocityViolation}

We first revisit the closed-form expression we derived in standard RT for the single scattering approximation in (\ref{e:1_scatter_R})
for radiances escaping the upper boundary.  We remarked that they have the reciprocity property that reversing $\Omvec_0$ (source) and $\Omvec$ (detector) in both sign and in order gives the same answer.

Here, we need to evaluate
\begin{equation}
I_1(0,\Omvec;\Omvec_0) =
\omega p_g(\Omvec\cdot\Omvec_0)
\int\limits_0^{\tau^\star} T_a(\tau^\prime/\mu_0) \,
|\Dot{T}_a(\tau^\prime/|\mu|)| \, \dif\tau^\prime/|\mu|.
\label{e:1_scatter_gRT_reflection}
\end{equation}
From there, (\ref{e:BRF_form}) yields for the BRF form
\begin{equation}
\begin{array}{l}
\frac{\pi}{\mu_0} I_1(0,\Omvec;\Omvec_0) = \pi \omega p_g(\Omvec\cdot\Omvec_0)
\times \frac{a}{\mu_0}
\left( \frac{\mu_0}{|\mu|} \right)^a
\left( 1-\frac{\mu_0}{|\mu|} \right)^{-2a} \\
\quad\quad\quad\quad\quad
\times \left[
 \mathrm{B}\left(2a,1-a;1-\frac{\mu_0}{|\mu|}\right)
-\mathrm{B}\left(2a,1-a;a\,\frac{1-\mu_0/|\mu|}{a+\tau^\star/|\mu|}\right)
\right],
\end{array}
\label{e:1_scatter_gR}
\end{equation}
with $-1 \le \mu < 0$ and $0 < \mu_0 \le 0$), and where we use the incomplete Euler Beta function: $\mathrm{B}(x,y;z) = \int_0^z t^{x-1} (1-t)^{y-1} \dif t$.

\begin{figure}[ht]
\begin{center}
\includegraphics[width=2.5in]{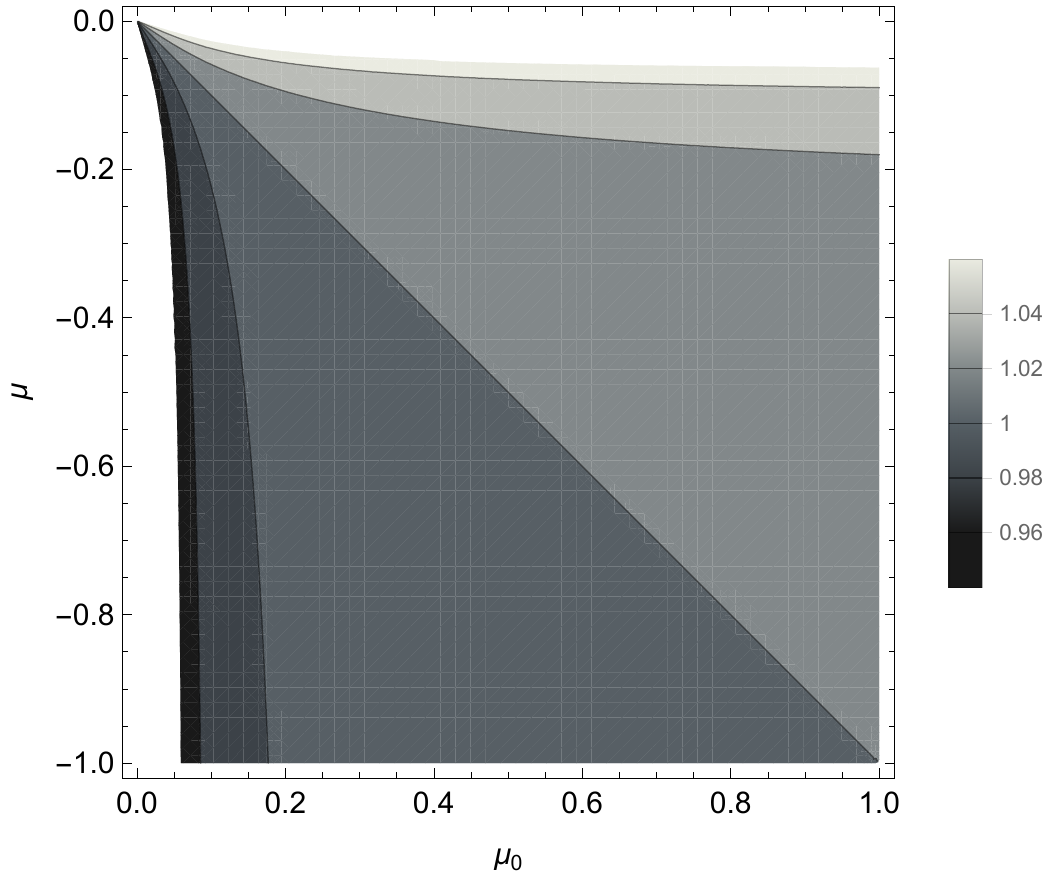}
\includegraphics[width=2.5in]{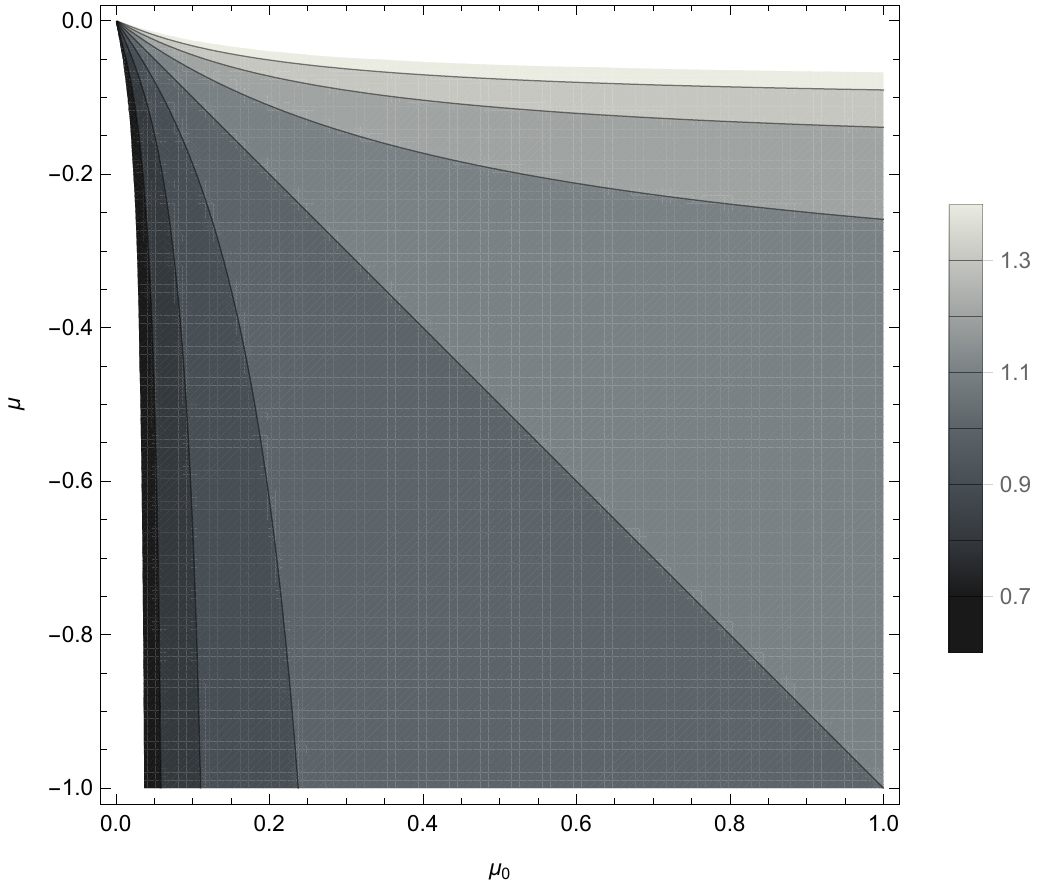}
\end{center}
\caption[NonRecip_gRT]
{\label{f:NonRecip_gRT}
Contour plots of $I_1(0,-\Omvec_0;-\Omvec)/|\mu| \div I_1(0,\Omvec;\Omvec_0)/\mu_0$ as functions of $\mu_0$ and $\mu$ for $\tau^\star = 0.1$ (a small value consistent with the adopted single scattering approximation) and $a=10$ (left), $a=1.2$ (right).}
\end{figure}

To demonstrate that this complex expression violates the reciprocity relation in (\ref{e:R_reciprocity}) and by how much, we have plotted in Fig.~\ref{f:NonRecip_gRT} the ratio of $I_1(0,-\Omvec_0;-\Omvec)/|\mu|$ to $I_1(0,\Omvec;\Omvec_0)/\mu_0$ for a small value of $\tau^\star$ compatible with the single scattering approximation used in (\ref{e:1_scatter_gRT_reflection}).  This ratio is independent of the SSA, $\omega$, and of azimuthal angle, $\phi$ (assuming $\phi_0 = 0$), in $d = 3$ since it appears only in the evaluation of the PF via $\Omvec_0\cdot\Omvec = \mu_0\mu+\sqrt{1-\mu_0^2}\sqrt{1-\mu^2}\cos\phi$.  As expected, the violation is stronger for smaller values of $a$ ($a$ = 1.2 and $a$ = 10 are displayed).

This violation of reciprocity is a desirable attribute of stochastic RT modeling at least in atmospheric applications.  It is indeed consistent with real-world satellite observations of reciprocity violation uncovered by DiGirolamo et al. \cite{DiGirolamo_EtAl_1998} in spatially variable cloud scenes inside a relatively broad field of view, and readily replicated with numerical Monte Carlo simulations.  These findings were soon explained theoretically by Leroy \cite{Leroy2001}.  This provides an element of validation of the new model and, by the same token, invalidates for atmospheric applications all models for RT in stochastic media based on either homogenization or linear mixing.

It is important to realize that this reciprocity violation is related ({\it i}) to the uniform illumination of the scene and ({\it ii}) to the spatial averaging that is inherent in the observations that the new model is designed to predict.  Indeed, at the scale of a collimated source at a single point in space and a collimated receiver aimed at another direction at another point in any medium, spatially variable or not, there is a fundamental principle of reciprocity as long as the PF has it, $p(\Omvec^\prime\to\Omvec) = p(-\Omvec\to-\Omvec^\prime)$, and Helmholtz's reciprocity principles will guaranty that property under most circumstances.  Starting from there, Case \cite{Case1957} showed that invariance under arbitrary horizontal translation  is also required to extend this (internal) ``Green's function'' reciprocity to Chandrasekhar's \cite{Chandra1950} (external) reciprocity relations for plane-parallel slabs.

\section{Conclusions \& Outlook}
\label{s:concl}

We have surveyed a still small but growing literature on radiative transfer (equivalently, mono-group linear transport) theory where there is no requirement for the direct transmission law---hence the propagation kernel---to be exponential in optical distance.  In particular, we gather the evidence from the atmospheric radiation and turbulence/cloud literatures that a better choice of transmission law \emph{on average} would have a power-law tail, at least for solar radiative transfer in large domains with a strong but unresolved variability of clouds and aerosols that is shaped by turbulent dynamics.  Long-range spatial correlations in the fluctuations of the extinction coefficient in the stochastic medium are essential to the emergence of power-law transmission laws, and such correlations are indeed omnipresent in turbulent media such as cloudy airmasses as well as entire cloud fields.

From there, we modified the integral form of the radiative transfer equation to accomodate such power-law kernels.  This leads to a \emph{generalized} linear transport theory parameterized by the power-law exponent.  This new model reverts to the standard one where exponential transmission prevails in the limit where the characteristic power-law exponent increases without bound.  In the new theory however, the physics dictate that there are two specific roles for the transmission function, which lead to different but related expressions.  There is no such formal distinction in standard transport theory.  However, when the origins of the exponentials are carefully scrutinized from a transport physics perspective, their different functionalities become apparent.

The new transport theory, possibly with some restrictions, is likely to be one instance of the new ``non-classical'' class of transport models investigated recently by Larsen and Vasques \cite{LarsenVasques11}.  These authors were primarily motivated by fundamental questions about neutron multiplication processes in pebble-bed nuclear reactors.  We do not anticipate long-range spatial correlations in these reactors so the relevant transmission laws are more likely to be modified exponentials such as found by Davis and Mineev-Weinstein \cite{DavisMineev11} in media with very high frequency fluctuations.

We presented a unified formulation for standard and generalized transport theory in $d = 1,2,3$ spatial dimensions and their associated direction spaces.  The present study first adds to previous ones the capability of a new deterministic computational scheme for solving the generalized linear transport equation, which does not have at present an integro-differential form, only an integral one.  We thus address the stochastic transport problem at hand, so far only in $d=1$, using a Markov chain formalism.  It is used to explore internal intensity and flux fields where numerical results shed light on questions of similarity and diffusion.  Diffusion theory and the space-angle similarity captured in the scaled or ``transport'' mean-free-path are exact in $d=1$ for standard transport---not so in generalized transport.  In $d>1$, diffusion is only an approximation applicable to opaque scattering media, and the associated similarity is only asymptotic (large optical thickness regimes).

New numerical simulations presented here in $d=2$ confirm and qualify the violation of similarly.  They also confirm previous predictions about the asymptotic scaling of diffuse transmission, which is anomalous or ``L\'evy-like'' if the characteristic exponent is less than 2.  L\'evy flights are now attracting considerable interest in laboratory as well as atmospheric optics \cite{BarthelemyEtAl2008}.  Finally, the generalized transport problem is solved in $d=3$ in the single scattering approximation.  This solution is used to highlight the violation of angular reciprocity in generalized radiative transfer.  This is yet another distinction between standard transport theory, including homogenization-based models for transport in stochastic media, and the new class of generalized transport models.  This non-reciprocity is in fact observed in the Earth's cloudy atmosphere using reflected sunlight, and is therefore a desirable attribute for stochastic transport of solar radiation.

A logical next step is to implement the Markov chain solution in $d = 3$.  Monte Carlo-based predictions of the angular patterns for radiance escaping the medium on the upper (obliquely illuminated) boundary are already available for the verification process.  In $d = 3$, there may be an interest in adding light polarization capability and linearizing the model with respect to the new spatial variability parameter $a$, the exponent that controls the power law tail of the direct transmission law.

Finally, we draw attention to a serendipitous development in the atmospheric radiation literature.  While our ongoing theoretical and computational work on unresolved/stochastic \emph{spatial} variability of the extinction coefficient in turbulent scattering media has lead to the parameterized class of power-law propagation kernels described herein, Conley and Collins \cite{ConleyCollins2011} have independently arrived at the very same power-law parameterization for the problem of unresolved \emph{spectral} variability of the absorption coefficient due to all manner of molecules in the Earth's atmosphere that might contribute to the thermal and solar radiative processes that build up the greenhouse effect.

This opens tantalizing questions about novel \emph{unified} formulations of the challenging problem of radiation transport in clumpy 3D scattering media that are permeated with spatially uniform but spectrally variable absorbing gases.  A first step in that direction is to assume that the scattering elements of the optical medium are in fact spatially uniform.  However, our generalized radiation transport model for multiple scattering based on power-law transmission between scatterings/absorptions still applies, and it can be invoked to capture the impact of purely spectral variability on the overall radiation transport from sources to sinks/detectors.

\newpage
\section{Appendix: Markov Chain Formalism for Generalized Radiative Transfer}
\label{s:append}
This page (35) is intentionally left blank, to be replaced by corresponding Appendix page.  N.B. arXiv users will find a PDF version of the Appendix as an ancillary file accessible from the abstract page.

\newpage
This page (36) is intentionally left blank, to be replaced by corresponding Appendix page.  N.B. arXiv users will find a PDF version of the Appendix as an ancillary file accessible from the abstract page.

\newpage
This page (37) is intentionally left blank, to be replaced by corresponding Appendix page.  N.B. arXiv users will find a PDF version of the Appendix as an ancillary file accessible from the abstract page.

\newpage
This page (38) is intentionally left blank, to be replaced by corresponding Appendix page.  N.B. arXiv users will find a PDF version of the Appendix as an ancillary file accessible from the abstract page.

\newpage
This page (39) is intentionally left blank, to be replaced by corresponding Appendix page.  N.B. arXiv users will find a PDF version of the Appendix as an ancillary file accessible from the abstract page.

\newpage

\section*{Acknowledgments}

The authors are thankful for sustained support from NASA's Radiation Sciences Programs managed by Hal Maring and Lucia Tsaoussi.  We also acknowledge many fruitful discussions with Howard Barker, Bob Cahalan, Bill Collins, Martin Frank, Lee Harrison, Alex Kostinski, Ed Larsen, Shaun Lovejoy, Alexander Marshak, Qilong Min, Klaus Pfeilsticker, Anil Prinja, Ray Shaw, and Bob West.  We also thank the Editors and two anonymous reviewers for many insightful comments about the submitted manuscript.  This research was carried out at the Jet Propulsion Laboratory, California Institute of Technology, under a contract with the National Aeronautics and Space Administration.

\textcolor{white}{\cite{Marchuk_etal80}}

\newpage

\bibliographystyle{unsrt}
\bibliography{TTSP_DavisXu}

Copyright 2014 California Institute of Technology. Government sponsorship acknowledged.

\end{document}